\documentclass[a4paper,11pt]{article}
\usepackage{jheppub} 
\usepackage{lineno}

\usepackage{tikz}

\allowdisplaybreaks[1]

\arxivnumber{2410.06727} 

\title{\boldmath Kramers-Wannier self-duality and non-invertible translation symmetry in quantum chains:~a wave-function perspective}

\author{Hua-Chen Zhang}
\author{and Germ\'an Sierra}
\affiliation{Instituto de F\'isica Te\'orica UAM/CSIC,\\
C/ Nicol\'as Cabrera 13-15, Cantoblanco, 28049 Madrid, Spain}

\emailAdd{huachen.zhang@ift.csic.es}
\emailAdd{german.sierra@csic.es}

\abstract{The Kramers-Wannier self-duality of critical quantum chains is examined from the perspective of model wave functions. We demonstrate, using the transverse-field Ising chain and the 3-state Potts chain as examples, that the symmetry operator for the Kramers-Wannier self-duality follows in a simple and direct way from a `generalised' translation symmetry of the model wave function in the anyonic fusion basis. This translation operation, in turn, comprises a sequence of \textit{F}-moves in the underlying fusion category. The symmetry operator thus obtained naturally admits the form of a matrix product operator and obeys non-invertible fusion rules. The findings reveal an intriguing connection between the (non-invertible) translation symmetry on the lattice and topological aspects of the conformal field theory describing the scaling limit.}


\begin{document}
\maketitle
\flushbottom

\section{Introduction}


Symmetry is an ancient yet timeless notion in physical science. Since the last few years, the scope of this concept has been expanded significantly, for which two of the most notable directions are known as the higher-form symmetries~\cite{gaiotto2015} and the non-invertible symmetries~\cite{brunner2015}, respectively. These developments, together with others, sometimes go under the overarching name of `generalised symmetries'. While the higher-form symmetries are relevant to physics in higher spacetime dimensions, the non-invertible symmetries are most thoroughly studied in (1+1)-dimensions [(1+1)D], where extensive investigations in the context of both continuum field theories~\cite{bhardwaj2018,chang2019,thorngren2019,thorngren2021,huang2021,huang2022,choi2024} and lattice models~\cite{aasen2016,belletête2023,ji2020,belletête2020,aasen2020,vanhove2022,inamura2022,lootens2023,lootens2024,eck2024,cao2023,seiberg2024a,sinha2024,seiberg2024b,okada2024,seifnashri2024,bhardwaj2024a,chatterjee2024,bhardwaj2024b,li2024,cao2024,pace2025} were carried out.

From the modern point of view, (generalised) symmetries in field theories are identified with~\emph{topological operators} that can be deformed through the spacetime without altering the correlation function (or partition function)~\cite{mcgreevy2023,schafernameki2023,bhardwaj2023,luo2024,shao2024}. Specifically, the topological operators pertinent to the (0-form) symmetries in (1+1)D are~\emph{topological line operators}. The latter terminology is often used interchangeably with `symmetry defects' or `topological defects', especially in the literature on continuum field theories. However, with the applications to quantum chains in mind, it is useful to distinguish between `spacelike' and `timelike' line operators. In particular, following refs.~\cite{seiberg2024a,seiberg2024b,sinha2024}, we reserve the phrase `symmetry defects' for the line operators stretching in the direction of (imaginary) time (along which the transfer matrix acts), and refer to those in the spatial direction simply as~\emph{symmetry operators}. For the case of quantum chains, the symmetry defects correspond to modifications of certain local terms in the Hamiltonian, whilst the symmetry operators act on the Hilbert space of the entire chain at a specific `time slice'. For~\emph{non-invertible} symmetries, the composition (or `fusion') of two topological operators possibly produces the direct sum of more than one topological operators. Clearly, these topological operators do not form a group; the appropriate mathematical structure modelling them turned out to be~\emph{fusion categories}.\footnote{By definition, the number of simple objects (namely, those that cannot be expressed as the sum of others) in a fusion category is finite. Nevertheless, many `good' properties of fusion categories are still valid if we do not assume this finiteness condition. The finiteness condition is violated, for example, in the case of a continuous global symmetry.} In this work, we shall use only a few bits from this subject, which will be reviewed at suitable places in the text; we refer the reader to the excellent textbooks~\cite{etingof2015,turaev2016,turaev2017} for systematic introductions. Non-invertible symmetries are therefore also known as categorical symmetries, which are in fact fairly ubiquitous in many branches of mathematics and physics, e.g., the 2D rational conformal field theory (CFT)~\cite{froehlich2004,froehlich2007,froehlich2010,carqueville2016,brunner2014,brunner2013}. There exist noteworthy methods to generate non-invertible symmetries, amongst which we would like to mention the so-called gauging (or orbifolding) of an `ordinary' discrete group symmetry~\cite{thorngren2021,li2023,shao2024}; the interface between the original and gauged theories becomes the line operator for a non-invertible symmetry when the theory is invariant under this gauging procedure.

The Kramers-Wannier (KW) duality transformation is a prototypical example of non-invertible transformations. Historically, this duality transformation was introduced as a mapping relating the partition functions of the 2D classical Ising model on the square lattice in the ordered and disordered phases, respectively~\cite{kramers1941}. The non-invertibility of KW transformation can be understood in the quantum-chain limit with the simple fact that the ground states in the ordered phase are two-fold degenerate, whereas one finds only one ground state in the disordered phase. At the critical point between the two phases, the partition function is self-dual and the KW transformation becomes a non-invertible~\emph{symmetry}. In fact, it was the self-duality combined with the uniqueness of the phase transition that enabled Kramers and Wannier to nail down the transition temperature of this model prior to Onsager's exact solution~\cite{onsager1944}. The structure of KW duality is familiar in the Ising and Potts~\cite{wu1982} models, and its formulation in terms of topological line operators has been generalised to a much larger class of 2D classical statistical models and the associated quantum chains~\cite{aasen2020,lootens2023,lootens2024}. Not surprisingly, the KW-type line operators also manifest themselves in the CFTs that describe the scaling limit of many of these models at their critical point. Following the approach mentioned at the end of the last paragraph, the KW-type self-duality of a 2D CFT arises from the invariance upon gauging a (non-anomalous) $\mathbb{Z}_n$ symmetry. Mathematically, the above situation is characterised by the $\mathbb{Z}_n$ Tambara-Yamagami category~\cite{tambara1998}, which is the extension of the $\mathbb{Z}_n$ group by a duality-like line. By definition, this fusion category consists of $(n+1)$ simple objects, where the duality-like line is the only non-invertible one, and the remaining $n$ lines fuse among themselves according to the $\mathbb{Z}_n$ group multiplication rules. 
In particular, the non-invertible lines in the cases $n=2$ and $n=3$ correspond to the KW self-duality of the Ising CFT and the $3$-state Potts CFT, respectively. More information on these CFTs and the associated fusion categories can be found in the text below.

The relation between the symmetries at the lattice level and those of the field theories describing the low-energy continuum limit, however, can usually be a subtle one. In refs.~\cite{seiberg2024a,seiberg2024b}, the authors performed a comprehensive study of various spacetime and internal symmetries of the Majorana chain, its bosonised counterpart (i.e., the quantum Ising chain), and the field theory in the continuum limit. One crucial point made by them is that the composition laws of the KW symmetry operator for the critical Ising chain get mixed with lattice translations, and hence do~\emph{not} form a fusion category.\footnote{The KW self-duality of the Ising CFT was described by the authors of refs.~\cite{seiberg2024a,seiberg2024b} as a non-invertible symmetry that~\emph{emanates} from the corresponding lattice symmetry.} Another instance is given in ref.~\cite{sinha2024}, where the authors found that the realisation of the Fibonacci topological line in the $3$-state Potts CFT becomes non-topological on the lattice. In the present work, we aim to investigate the related issues from another perspective, namely that of~\emph{model wave functions}, with a focus on the KW self-duality, using the quantum Ising chain and the 3-state Potts chain as concrete examples.

The construction of model wave functions has played an important role in our understanding of the physics of quantum many-body systems, such as the fractional quantum Hall effects~\cite{tsui1982,laughlin1983}. The~\emph{exact} eigenstates of a many-body Hamiltonian are typically unknown or have extremely complicated forms (with remarkable exceptions; see below). The power of model wave functions lies in that they capture the relevant properties of the ground state(s) or low-lying excited states while admitting relatively simple forms and being analytically or numerically tractable. The seminal work by Moore and Read~\cite{moore1991} opened up new horizons in the research on model wave functions. The Pfaffian wave function they proposed accommodates anyonic excitations with non-Abelian statistics and serves as a candidate for the fractional quantum Hall state at filling fraction $5/2$~\cite{willett1987}, making the latter potentially useful in designing topological quantum devices~\cite{nayak2008}. On the methodological side, moreover, they ingeniously established the connection between the model wave functions and certain chiral correlators (also known as~\emph{conformal blocks}) in a CFT, which makes it possible to construct and study model wave functions with techniques from CFT. This approach was later adapted to lattice quantum systems in both one~\cite{cirac2010} and two~\cite{nielsen2012} spatial dimensions. Model wave functions constructed using conformal blocks inherit elegant properties from the CFT, especially concerning the (non-invertible) symmetries. This stems largely from the common categorical structure shared by the topological lines and the `chiral data' of a 2D rational CFT, which we shall briefly review in the next subsection.

\subsection{Chiral data of a 2D rational CFT: a cursory review}

\begin{figure}[htbp]
\centering
\includegraphics[width=\textwidth]{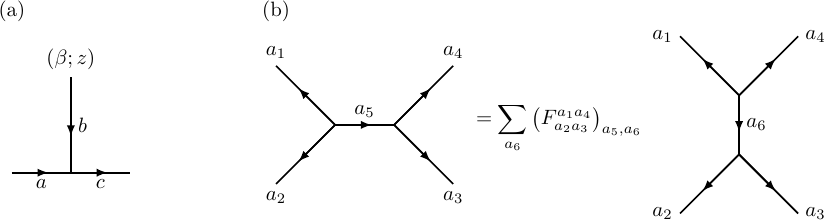}
\caption{(a) Diagrammatic representation of the chiral vertex operator $\Phi_{ab}^{c}(\beta;z)$. (b) The definition of the $F$-move in terms of four-point conformal blocks. \label{fig:CVO-and-F-move}}
\end{figure}

The symmetry algebra of a 2D CFT~\cite{belavin1984,francesco1997} has the form $\mathcal{A}\times\Bar{\mathcal{A}}$, where $\mathcal{A}$ and $\Bar{\mathcal{A}}$ are called the left- and right-moving chiral algebras, respectively. By definition, both $\mathcal{A}$ and $\Bar{\mathcal{A}}$ must contain the Virasoro algebra. The state space of the theory typically admits the decomposition into a direct sum of tensor products as follows:
\begin{equation}
\label{eq:state-space-decomposition}
    \mathcal{H} = \bigoplus_{(a,\Bar{a})} M_{a,\Bar{a}} \mathcal{H}_{a} \otimes \mathcal{H}_{\Bar{a}},
\end{equation}
where $\mathcal{H}_{a}$ (resp. $\mathcal{H}_{\Bar{a}}$) is an irreducible highest-weight representation of $\mathcal{A}$ (resp. $\Bar{\mathcal{A}}$), and the non-negative integer $M_{a,\Bar{a}}$ is the multiplicity for the combination $\mathcal{H}_{a} \otimes \mathcal{H}_{\Bar{a}}$ to occur. $(a,\Bar{a})$ belongs to some index set, which is finite for a~\emph{rational} CFT. The left- or right-moving irrep with the vacuum state as highest-weight vector is denoted as $\mathcal{H}_{\boldsymbol{1}}$; we require that $M_{\boldsymbol{1},\boldsymbol{1}} = 1$. We note that the monodromy invariance of physical correlators, the associativity of operator product expansions and the modular invariance of the torus partition function impose highly non-trivial constraints on the possible combinations of left- and right-moving irreps~\cite{moore1988,moore1989,moore1990}. Here, however, let us focus on the irreps of one of the chiral algebras, say $\mathcal{A}$. It turns out that a fusion product can be defined for these irreps~\cite{gaberdiel1994}; if $\mathcal{H}_{c}$ appears in the fusion product of $\mathcal{H}_{a}$ and $\mathcal{H}_{b}$, there exists a~\emph{chiral vertex operator} (CVO)\footnote{For simplicity, we assume that there is only one fusion channel for $\mathcal{H}_{c}$ in the fusion product of $\mathcal{H}_{a}$ and $\mathcal{H}_{b}$.}
\begin{equation}
    \Phi_{ab}^{c}(\beta;z):~\mathcal{H}_{a} \rightarrow \mathcal{H}_{c}
\end{equation}
for each state $\beta \in \mathcal{H}_{b}$; $z$ is the coordinate of the point on the `worldsheet' at which the CVO is inserted. When $\beta$ is the highest-weight state, $\Phi_{ab}^{c}(\beta;z)$ is a~\emph{primary} CVO. The diagrammatic representation of the CVO is shown in figure~\ref{fig:CVO-and-F-move}(a), where each line is oriented with an arrow and labelled by an irrep (say $\mathcal{H}_{a}$); reversing the orientation amounts to replacing $\mathcal{H}_{a}$ by its conjugation, which is defined as the (unique) irrep whose fusion product with $\mathcal{H}_{a}$ contains $\mathcal{H}_{\boldsymbol{1}}$.

Conformal blocks are nothing but correlators of the CVOs. In fact, the CVO $\Phi_{ab}^{c}(\beta;z)$ itself can be regarded as a three-point conformal block if one has specified a state in each of the irreps $\mathcal{H}_{a}$ and $\mathcal{H}_{c}$. Proceeding to the four-point case, one finds an important duality property known as the $F$-move, which corresponds to the basis transformation shown in figure~\ref{fig:CVO-and-F-move}(b), where the transformation matrix $F^{a_{1}a_{4}}_{a_{2}a_{3}}$ is called an $F$-symbol (here, our notation follows refs.~\cite{chang2019,thorngren2021}). This relation is interpreted as an identity for the linear mappings acting in the tensor products of the corresponding irreps. The $F$-moves will be the main workhorse in our construction of KW symmetry operators below. In addition to the $F$-moves, the braiding, or $B$-moves, can be defined by the monodromy under analytical continuations for the coordinates of the CVOs. The pentagon and hexagon identities arise as consistency constraints in the consideration of five-point blocks. These constraints, also collectively known as the Moore-Seiberg polynomial equations~\cite{moore1988}, endow the irreps of $\mathcal{A}$ the structure of a~\emph{modular tensor category}. This category is sometimes referred to as the~\emph{chiral data} and, roughly speaking, reduces to a fusion category if one `forgets the braiding'. We conclude that the chiral data of a 2D rational CFT come with a fusion category, as the topological line operators do.

\subsection{Main results and the structure of the paper}

The key property of the model wave functions that we exploit in this work is what we will call the `half-step translation symmetry', which is defined by a sequence of $F$-moves in the Tambara-Yamagami fusion category and admits the meaning of `translation by half a site'.

We start gently with the quantum Ising chain in section~\ref{sec:Ising}. After recapitulating some basic facts about the Ising CFT, including the topological lines in it, we write down the model wave function for the critical Ising chain in the `fusion tree' basis of the $\mathbb{Z}_2$ Tambara-Yamagami category, which is constructed as a conformal block of certain primary CVOs. The operator implementing the half-step translation is then naturally expressed in terms of the $F$-symbols. We show that this operator is precisely the symmetry operator for the KW self-duality of the critical Ising chain; it can be brought into the form of a matrix product operator with periodic boundary condition and obeys non-invertible composition laws mixing with lattice translations. These results have been obtained in ref.~\cite{seiberg2024b} by a different method; nonetheless, we believe that re-deriving them for this simplest case is best suited for illustrating our approach, which applies to more complicated models.

Model wave functions based on CFTs make the definition of the half-step translation symmetry natural and intuitive. Nevertheless, this definition does~\emph{not} require the model wave function to be in the form of a conformal block (or its explicit expression to be known). In section~\ref{sec:3-state-Potts}, we will turn to the $\mathbb{Z}_3$ case, where the model wave function is no longer written as a conformal block in the corresponding CFT, yet the half-step translation operator can still be defined parallel to the Ising case. The critical $3$-state Potts chain is obtained as a Hamiltonian that commutes with this operator, which we identify to be the KW symmetry operator of the former. This expression for the symmetry operator is new to our knowledge. The composition laws involving this symmetry operator are derived and compared to the fusion rules in the $\mathbb{Z}_3$ Tambara-Yamagami category. The simplicity of the derivations in this section demonstrates that the approach we proposed for the Ising model generalises straightforwardly to more complicated models.

We must point out that the recognition of the KW duality transformation as `translating by half a site' has been made by other authors. There are, however, important differences between our perspective and those in the previous works. In section~\ref{sec:relationship}, the connections and differences between our approach and other existing ones are discussed. We summarise our findings and give an outlook on some related open problems in section~\ref{sec:summary}. Some details omitted from the main text are collected in three short appendices.

\section{Kramers-Wannier symmetry operator for the quantum Ising chain}
\label{sec:Ising}

\subsection{Ising CFT and fusion category}

The critical quantum Ising chain in a transverse magnetic field, with Hamiltonian
\begin{equation}
\label{eq:Ising-Hamiltonian}
    \widehat{H} = -\sum_{l=1}^{L-1} X_{l}X_{l+1} - X_{L}X_{1} - \sum_{l=1}^{L} Z_{l}
\end{equation}
on a lattice of $L$ sites with periodic boundary condition, is obtained from an anisotropic limit of the transfer matrix for the critical 2D classical Ising model~\cite{fradkin1978,kogut1979}. The $X_{l}$ and $Z_{l}$ are Pauli operators, for which the matrix representation in the standard basis is
\begin{equation}
    X_{l} = \left( \begin{array}{cc}
       0  &~ 1 \\
        1 &~ 0
    \end{array} \right)_{l}, \quad Z_{l} = \left( \begin{array}{cc}
       1  &~ 0 \\
        0 &~ -1
    \end{array} \right)_{l}.
\end{equation}
The CFT describing the scaling limit of this critical point is identified as the Virasoro minimal model $\mathcal{M}(3,4)$ with central charge $1/2$~\cite{belavin1984}. For the minimal models, each of the chiral algebras $\mathcal{A}$ and $\Bar{\mathcal{A}}$ is isomorphic to the Virasoro algebra $Vir$ itself. In the case of ${Vir}(3,4)$, there are three irreps; we use the same symbols $a = \boldsymbol{1}, \chi, \sigma$ to label these irreps and the corresponding primary CVOs, which are the identity, the chiral Majorana field and the chiral part of the spin field, with conformal weights $0$, $1/2$ and $1/16$, respectively. The non-trivial fusion rules read
\begin{equation}
\label{eq:Ising-primaries-fusion-rules}
    \chi \times \chi = \boldsymbol{1}, \quad \chi \times \sigma = \sigma, \quad \sigma \times \sigma = \boldsymbol{1} + \chi.
\end{equation}

The Ising CFT is an example of~\emph{diagonal} rational CFTs, for which one has $M_{a,\Bar{a}} = \delta_{a,\Bar{a}}$ in~\eqref{eq:state-space-decomposition}. In these theories, there are a family of topological line operators, called the~\emph{Verlinde lines}~\cite{verlinde1988,petkova2001}, which commute with both chiral algebras. As an example of the so-called modular bootstrap, the Verlinde lines are in one-to-one correspondence with the irreps of the chiral algebras; this is not coincidental but has deep roots in the modular invariance~\cite{cardy1986}. The Verlinde lines corresponding to $\boldsymbol{1}$, $\chi$ and $\sigma$ are the trivial line, the spin-flip line and the KW duality line, denoted as $\mathcal{L}_{\boldsymbol{1}}$, $\mathcal{L}_{\eta}$ and $\mathcal{L}_{\mathcal{N}}$, respectively. Their fusion rules are identical to those of the chiral primaries,~\eqref{eq:Ising-primaries-fusion-rules}:
\begin{equation}
    \mathcal{L}_{\eta}^{2} = \mathcal{L}_{\boldsymbol{1}}, \quad \mathcal{L}_{\eta} \mathcal{L}_{\mathcal{N}} = \mathcal{L}_{\mathcal{N}} \mathcal{L}_{\eta} = \mathcal{L}_{\mathcal{N}}, \quad \mathcal{L}_{\mathcal{N}}^{2} = \mathcal{L}_{\boldsymbol{1}} + \mathcal{L}_{\eta}.
\end{equation}
In fact, the fusion category formed by the Verlinde lines or chiral primaries in the Ising CFT is nothing but the $\mathbb{Z}_2$ Tambara-Yamagami category\footnote{\label{ftnt:FS}More precisely, it corresponds to the $\mathbb{Z}_2$ Tambara-Yamagami category with a specific choice of a sign known as the Frobenius-Schur indicator~\cite{tambara1998,simon2022}, which is associated with the fact that $\mathcal{L}_{\mathcal{N}}$ (or $\sigma$) is self-conjugate. However, as noted in ref.~\cite{seiberg2024b}, since the adjoint of the~\emph{lattice} KW symmetry operator involves translation and is~\emph{not} the same as the KW operator itself, this subtlety is immaterial in the case we shall be focusing on. The same applies to the $\mathbb{Z}_3$ case.}, as $\{ \mathcal{L}_{\boldsymbol{1}}, \mathcal{L}_{\eta} \}$ generate the group $\mathbb{Z}_2$. The remaining data of this fusion category, namely its $F$-symbols, are listed below for later utility~\cite{chang2019}:
\begin{equation}
\label{eq:F-symbols-Z2}
    \left( F^{\chi\sigma}_{\sigma\chi} \right)_{\sigma,\sigma} = -1,
    \quad
    F^{\sigma\sigma}_{\sigma\sigma} = \frac{1}{\sqrt{2}} \left( \begin{array}{cc}
       1  &~ 1 \\
        1 &~ -1
    \end{array} \right),
\end{equation}
and all other $F$-moves are trivial. Note that the rows and columns of $F^{\sigma\sigma}_{\sigma\sigma}$ are arranged in the order $(\boldsymbol{1}, \chi)$.

\subsection{Model wave function and its half-step translation symmetry}
\label{subsec:Ising-model-wave-function}

A model wave function for the critical ground state of the quantum Ising chain~\eqref{eq:Ising-Hamiltonian}, based on the conformal blocks in the Ising CFT, was proposed in ref.~\cite{montes2017a}. As it turned out~\cite{montes2017b}, this model state is in fact the~\emph{exact} ground state of~\eqref{eq:Ising-Hamiltonian}. This fact is of interest in its own right, and we shall come back to it in section~\ref{sec:summary}; as the very simplest example, the ground state of the critical Ising chain with $2$ sites is explicitly compared to the $4$-point conformal block in appendix~\ref{append:O}. Nevertheless, it will not be used in our derivation of the KW symmetry operator below; what is important to us is the so-called half-step translation symmetry of this model wave function. In this subsection, we introduce the construction of the model wave function and define its half-step translation symmetry.

\begin{figure}[htbp]
\centering
\includegraphics[width=0.618\textwidth]{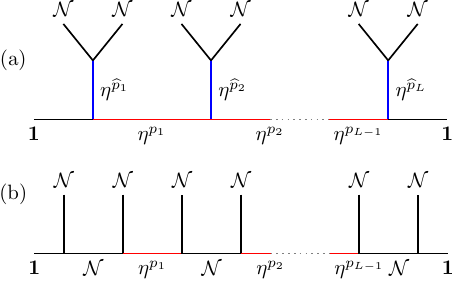}
\caption{The `pairwise fusion' (a) and multiperipheral (b) representations of the fusion tree in the $\mathbb{Z}_2$ Tambara-Yamagami category. Here, $\eta^{p_{l}} = \boldsymbol{1}$ for $p_{l} = 0$ and $\eta^{p_{l}} = \eta$ for $p_{l} = 1$; the same applies to $\eta^{\widehat{p}_{l}}$. \label{fig:Ising-fusion-trees}}
\end{figure}

The basis in which the model wave function was written down is defined by certain `fusion trees' from the $\mathbb{Z}_2$ Tambara-Yamagami category. This setup resembles that used for the restricted solid-on-solid (RSOS) or interaction round-a-face (IRF) models~\cite{andrews1984,pasquier1987,gomez1996}, the anyonic chains~\cite{feiguin2007,gils2013}, and the models proposed in ref.~\cite{aasen2020}.\footnote{The Ising and $3$-state Potts models can be obtained from the RSOS construction based on the Dynkin diagrams of the Lie algebras $\mathfrak{su}(4)$ and $\mathfrak{so}(8)$, respectively. Note, however, that the Hilbert space of some models from the RSOS/IRF or anyonic chain constructions does not admit factorisation into the tensor product of local ones. An example is given by the anyonic chains proposed in refs.~\cite{feiguin2007,gils2013}, where the Hilbert space is not of the tensor-product form, and the non-invertible symmetry (termed `topological symmetry') does not involve lattice translations.} A basis for the Hilbert (sub)space of~\eqref{eq:Ising-Hamiltonian}, $\{ | \widehat{p}_{1}, \ldots, \widehat{p}_{L} \rangle \}$ with the height variable $\widehat{p}_{l} = 0, 1$ and
\begin{equation}
    \quad X_{l} | \widehat{p}_{l} \rangle = | 1 - \widehat{p}_{l} \rangle, \quad Z_{l} | \widehat{p}_{l} \rangle = (-1)^{\widehat{p}_{l}} | \widehat{p}_{l} \rangle, \quad l = 1, \ldots, L,
\end{equation}
is encoded in the `pairwise fusion' representation of the fusion tree that is shown in figure~\ref{fig:Ising-fusion-trees}(a), where $\eta^{\widehat{p}_{l}} = \boldsymbol{1}$ for $\widehat{p}_{l} = 0$ and $\eta^{\widehat{p}_{l}} = \eta$ for $\widehat{p}_{l} = 1$. Note that $\widehat{p}_{1}, \ldots, \widehat{p}_{L}$ are not independent as $\sum_{l=1}^{L} \widehat{p}_{l} = 0~(\mathrm{mod}~2)$. Thus, $\{ | \widehat{p}_{1}, \ldots, \widehat{p}_{L} \rangle \}$ forms a basis for the Neveu-Schwarz (NS) sector, which is the $2^{L-1}$-dimensional subspace, corresponding to the projector
\begin{equation}
    P_{+} = \frac{1}{2} \left( 1 + \prod_{l=1}^{L} Z_{l} \right)~\text{with}~[\widehat{H}, P_{+}] = 0,
\end{equation}
of the total $2^{L}$-dimensional Hilbert space of the model~\eqref{eq:Ising-Hamiltonian}. It is known that the ground state of~\eqref{eq:Ising-Hamiltonian} resides in the NS sector. Alternatively, one can use $\{ | p_{1}, \ldots, p_{L-1} \rangle \}$ as a basis for this subspace, where $p_{1}, \ldots, p_{L-1}$ ($=0, 1$)~\emph{are} independent and the $\widehat{p}_{l}$ are related to them via
\begin{equation}
    \widehat{p}_{l} = p_{l} - p_{l-1}~(\mathrm{mod}~2),~l = 1, \ldots, L,~\text{where}~p_{0}  = p_{L} \equiv 0.
\end{equation}
In fact, the fusion trees in the `pairwise fusion' representation can be mapped to those in the more familiar multiperipheral representation [figure~\ref{fig:Ising-fusion-trees}(b)] through a sequence of trivial $F$-moves. The model state is taken to be
\begin{subequations}
\label{eq:Ising-GS-Ansatz}
\begin{equation}
    | \Psi \rangle = \sum_{\mathbf{p}} \mathcal{F}_{\mathbf{p}}(z_{1}, \ldots, z_{N}) | \widehat{p}_{1}, \ldots, \widehat{p}_{L} \rangle,
\end{equation}
where
\begin{eqnarray}
\label{eq:Ising-CB}
    \mathcal{F}_{\mathbf{p}}(z_{1}, \ldots, z_{N}) &=& \langle \sigma(z_{1}) \cdots \sigma(z_{N}) \rangle_{\mathbf{p}} \nonumber \\
    &\equiv&
    \begin{tikzpicture}[baseline,vertex/.style={anchor=base,
        circle,fill=black!25,minimum size=18pt,inner sep=2pt}]
\draw[black, thick] (0,0-0.5) -- (0.5,0-0.5);
\draw[black, thick] (0.5,1-0.5) -- (0.5,0-0.5);
\draw[black, thick] (0.5,0-0.5) -- (1.5,0-0.5);
\draw[black, thick] (1.5,1-0.5) -- (1.5,0-0.5);
\draw[red, thick] (1.5,0-0.5) -- (2.5,0-0.5);
\draw[black, thick] (2.5,0-0.5) -- (2.5,1-0.5);
\draw[black, thick] (2.5,0-0.5) -- (3.5,0-0.5);
\draw[black, thick] (3.5,1-0.5) -- (3.5,0-0.5);
\draw[red, thick] (3.5,0-0.5) -- (4,0-0.5);
\draw[dotted, gray, thick] (4,0-0.5) -- (5,0-0.5);
\draw[red, thick] (5,0-0.5) -- (5.5,0-0.5);
\draw[black, thick] (5.5,1-0.5) -- (5.5,0-0.5);
\draw[black, thick] (5.5,0-0.5) -- (6.5,0-0.5);
\draw[black, thick] (6.5,0-0.5) -- (6.5,1-0.5);
\draw[black, thick] (6.5,0-0.5) -- (7,0-0.5);

\draw (0,0-0.5) node[anchor = north]{$\boldsymbol{1}$};
\draw (0.5,1-0.5) node[anchor = south]{$\sigma(z_{1})$};
\draw (1,0-0.5) node[anchor = north]{$\sigma$};
\draw (1.5,1-0.5) node[anchor = south]{$\sigma(z_{2})$};
\draw (2,0-0.5) node[anchor = north]{$a_{1}$};

\draw (2.5,1-0.5) node[anchor = south]{$\sigma(z_{3})$};
\draw (3,0-0.5) node[anchor = north]{$\sigma$};
\draw (3.5,1-0.5) node[anchor = south]{$\sigma(z_{4})$};
\draw (4,0-0.5) node[anchor = north]{$a_{2}$};

\draw (5.5,0-0.5) node[anchor = north]{$a_{L-1}$};
\draw (5.5,1-0.5) node[anchor = south]{$\sigma(z_{N-1})$};
\draw (6,0-0.5) node[anchor = north]{$\quad\sigma$};
\draw (6.5,1-0.5) node[anchor = south]{$\quad\sigma(z_{N})$};
\draw (7,0-0.5) node[anchor = north]{$\boldsymbol{1}$};

\end{tikzpicture}
\end{eqnarray}
\end{subequations}
is the $N$($=2L$)-point conformal block of the chiral primary $\sigma$ in the Ising CFT with $\mathbf{p} \equiv (p_{1}, \ldots, p_{L-1})$ labelling the internal fusion channels by the identification $a_{l} = \boldsymbol{1}$ (resp. $a_{l} = \chi$) for $p_{l} = 0$ (resp. $p_{l} = 1$). We notice that it is not necessary to put arrows on the lines in these diagrams, since all simple objects in the $\mathbb{Z}_2$ Tambara-Yamagami category are self-conjugate.

The coordinates $z_{1}, \ldots, z_{N} \in \mathbb{C}$ of the chiral primaries on the `worldsheet', which we choose to be the complex plane, are parameters of the model wave function. When $z_{1}, \ldots, z_{N}$ are distributed uniformly on a circle [e.g. the unit circle, for which $z_{j} = \mathrm{e}^{2\pi\mathrm{i}(j-1)/N}$], this model wave function is invariant under the operation of a `generalised' translation, which comprises a sequence of $F$-moves as follows:
\begin{eqnarray}
    &&\mathcal{F}_{\mathbf{p}}(z_{1}, \ldots, z_{N}) =
\begin{tikzpicture}[baseline,vertex/.style={anchor=base,
        circle,fill=black!25,minimum size=18pt,inner sep=2pt}]

\draw[black, thick] (0,0-0.5) -- (0.5,0-0.5);
\draw[black, thick] (0.5,1-0.5) -- (0.5,0-0.5);
\draw[red, thick] (0.5,0-0.5) -- (1.5,0-0.5);
\draw[black, thick] (1.5,1-0.5) -- (1.5,0-0.5);
\draw[black, thick] (1.5,0-0.5) -- (2.5,0-0.5);
\draw[black, thick] (2.5,0-0.5) -- (2.5,1-0.5);
\draw[red, thick] (2.5,0-0.5) -- (3.5,0-0.5);
\draw[black, thick] (3.5,1-0.5) -- (3.5,0-0.5);
\draw[black, thick] (3.5,0-0.5) -- (4,0-0.5);
\draw[dotted, gray, thick] (4,0-0.5) -- (5,0-0.5);
\draw[red, thick] (5,0-0.5) -- (5.5,0-0.5);
\draw[black, thick] (5.5,1-0.5) -- (5.5,0-0.5);
\draw[black, thick] (5.5,0-0.5) -- (6,0-0.5);

\draw (0,0-0.5) node[anchor = east]{$\sigma(z_{1})$};
\draw (0.5,1-0.5) node[anchor = south]{$\sigma(z_{2})$};
\draw (1,0-0.5) node[anchor = north]{$a_{1}$};
\draw (1.5,1-0.5) node[anchor = south]{$\sigma(z_{3})$};
\draw (2,0-0.5) node[anchor = north]{$\sigma$};

\draw (2.5,1-0.5) node[anchor = south]{$\sigma(z_{4})$};
\draw (3,0-0.5) node[anchor = north]{$a_{2}$};
\draw (3.5,1-0.5) node[anchor = south]{$\sigma(z_{5})$};
\draw (4,0-0.5) node[anchor = north]{$\sigma$};

\draw (5.5,0-0.5) node[anchor = north]{$a_{L-1}$};
\draw (5.5,1-0.5) node[anchor = south]{$\sigma(z_{N-1})$};
\draw (6,0-0.5) node[anchor = west]{$\sigma(z_{N})$};

\end{tikzpicture} \nonumber \\
&=& \sum_{a^{\prime}_{1}} \left( F^{\sigma\sigma}_{\sigma\sigma} \right)_{a_{1},a^{\prime}_{1}}
\begin{tikzpicture}[baseline,vertex/.style={anchor=base,
        circle,fill=black!25,minimum size=18pt,inner sep=2pt}]

\draw[black, thick] (1,0-0.5) -- (1.5,0-0.5);
\draw[black, thick] (1,0) -- (1.5,0);

\draw[black, thick] (1.5,1-0.5) -- (1.5,0-0.5);
\draw[orange, thick] (1.5,-0.5) -- (1.5,0);
\draw[black, thick] (1.5,0-0.5) -- (2.5,0-0.5);
\draw[black, thick] (2.5,0-0.5) -- (2.5,1-0.5);
\draw[red, thick] (2.5,0-0.5) -- (3.5,0-0.5);
\draw[black, thick] (3.5,1-0.5) -- (3.5,0-0.5);
\draw[black, thick] (3.5,0-0.5) -- (4,0-0.5);
\draw[dotted, gray, thick] (4,0-0.5) -- (5,0-0.5);
\draw[red, thick] (5,0-0.5) -- (5.5,0-0.5);
\draw[black, thick] (5.5,1-0.5) -- (5.5,0-0.5);
\draw[black, thick] (5.5,0-0.5) -- (6,0-0.5);

\draw (1,0-0.5) node[anchor = east]{$\sigma(z_{1})$};
\draw (1,0) node[anchor = east]{$\sigma(z_{2})$};
\draw (1.5,0-0.25) node[anchor = west]{$a^{\prime}_{1}$};
\draw (1.5,1-0.5) node[anchor = south]{$\sigma(z_{3})$};
\draw (2,0-0.5) node[anchor = north]{$\sigma$};

\draw (2.5,1-0.5) node[anchor = south]{$\sigma(z_{4})$};
\draw (3,0-0.5) node[anchor = north]{$a_{2}$};
\draw (3.5,1-0.5) node[anchor = south]{$\sigma(z_{5})$};
\draw (4,0-0.5) node[anchor = north]{$\sigma$};

\draw (5.5,0-0.5) node[anchor = north]{$a_{L-1}$};
\draw (5.5,1-0.5) node[anchor = south]{$\sigma(z_{N-1})$};
\draw (6,0-0.5) node[anchor = west]{$\sigma(z_{N})$};

\end{tikzpicture} \nonumber \\
&=& \sum_{a^{\prime}_{1}} \left( F^{\sigma\sigma}_{\sigma\sigma} \right)_{a_{1},a^{\prime}_{1}} \left( F^{a^{\prime}_{1} \sigma}_{\sigma a_{2}} \right)_{\sigma,\sigma}
\begin{tikzpicture}[baseline,vertex/.style={anchor=base,
        circle,fill=black!25,minimum size=18pt,inner sep=2pt}]

\draw[black, thick] (1,0-0.5) -- (1.5,0-0.5);
\draw[black, thick] (1,0) -- (1.5,0);

\draw[black, thick] (1.5,1-0.5) -- (1.5,0);
\draw[orange, thick] (1.5,0) -- (2.5,0);
\draw[black, thick] (1.5,0-0.5) -- (2.5,0-0.5);
\draw[black, thick] (2.5,0-0.5) -- (2.5,1-0.5);
\draw[red, thick] (2.5,0-0.5) -- (3.5,0-0.5);
\draw[black, thick] (3.5,1-0.5) -- (3.5,0-0.5);
\draw[black, thick] (3.5,0-0.5) -- (4,0-0.5);
\draw[dotted, gray, thick] (4,0-0.5) -- (5,0-0.5);
\draw[red, thick] (5,0-0.5) -- (5.5,0-0.5);
\draw[black, thick] (5.5,1-0.5) -- (5.5,0-0.5);
\draw[black, thick] (5.5,0-0.5) -- (6,0-0.5);

\draw (1,0-0.5) node[anchor = east]{$\sigma(z_{1})$};
\draw (1,0) node[anchor = east]{$\sigma(z_{2})$};
\draw (2,0) node[anchor = south]{$a^{\prime}_{1}$};
\draw (1.5,1-0.5) node[anchor = south]{$\sigma(z_{3})$};
\draw (2.5,-0.25) node[anchor = west]{$\sigma$};

\draw (2.5,1-0.5) node[anchor = south]{$\sigma(z_{4})$};
\draw (3,0-0.5) node[anchor = north]{$a_{2}$};
\draw (3.5,1-0.5) node[anchor = south]{$\sigma(z_{5})$};
\draw (4,0-0.5) node[anchor = north]{$\sigma$};

\draw (5.5,0-0.5) node[anchor = north]{$a_{L-1}$};
\draw (5.5,1-0.5) node[anchor = south]{$\sigma(z_{N-1})$};
\draw (6,0-0.5) node[anchor = west]{$\sigma(z_{N})$};

\end{tikzpicture} \nonumber \\
&=& \sum_{a^{\prime}_{1} a^{\prime}_{2}} \left( F^{\sigma\sigma}_{\sigma\sigma} \right)_{a_{1},a^{\prime}_{1}} \left( F^{a^{\prime}_{1} \sigma}_{\sigma a_{2}} \right)_{\sigma,\sigma} \left( F^{\sigma\sigma}_{\sigma\sigma} \right)_{a_{2},a^{\prime}_{2}}
\begin{tikzpicture}[baseline,vertex/.style={anchor=base,
        circle,fill=black!25,minimum size=18pt,inner sep=2pt}]

\draw[black, thick] (1,0-0.5) -- (3.5,0-0.5);
\draw[black, thick] (1,0) -- (1.5,0);
\draw[black, thick] (1.5,1-0.5) -- (1.5,0);
\draw[orange, thick] (1.5,0) -- (2.5,0);
\draw[black, thick] (2.5,0) -- (3.5,0);
\draw[black, thick] (2.5,0.5-0.5) -- (2.5,1-0.5);
\draw[black, thick] (3.5,1-0.5) -- (3.5,0-0.5);
\draw[orange, thick] (3.5,0.5-0.5) -- (3.5,0-0.5);
\draw[black, thick] (3.5,0-0.5) -- (4,0-0.5);
\draw[dotted, gray, thick] (4,0-0.5) -- (5,0-0.5);
\draw[red, thick] (5,0-0.5) -- (5.5,0-0.5);
\draw[black, thick] (5.5,1-0.5) -- (5.5,0-0.5);
\draw[black, thick] (5.5,0-0.5) -- (6,0-0.5);

\draw (1,0-0.5) node[anchor = east]{$\sigma(z_{1})$};
\draw (1,0) node[anchor = east]{$\sigma(z_{2})$};
\draw (2,0) node[anchor = south]{$a^{\prime}_{1}$};
\draw (1.5,1-0.5) node[anchor = south]{$\sigma(z_{3})$};
\draw (3,0) node[anchor = south]{$\sigma$};

\draw (2.5,1-0.5) node[anchor = south]{$\sigma(z_{4})$};
\draw (3.5,0-0.25) node[anchor = west]{$a^{\prime}_{2}$};
\draw (3.5,1-0.5) node[anchor = south]{$\sigma(z_{5})$};
\draw (4,0-0.5) node[anchor = north]{$\sigma$};

\draw (5.5,0-0.5) node[anchor = north]{$a_{L-1}$};
\draw (5.5,1-0.5) node[anchor = south]{$\sigma(z_{N-1})$};
\draw (6,0-0.5) node[anchor = west]{$\sigma(z_{N})$};

\end{tikzpicture} \nonumber \\
&=& \ldots \ldots \nonumber \\
&=& \sum_{a^{\prime}_{1} \ldots a^{\prime}_{L-1}} \left( F^{\sigma\sigma}_{\sigma\sigma} \right)_{a_{1},a^{\prime}_{1}} \left( F^{a^{\prime}_{1} \sigma}_{\sigma a_{2}} \right)_{\sigma,\sigma} \left( F^{\sigma\sigma}_{\sigma\sigma} \right)_{a_{2},a^{\prime}_{2}} \cdots \left( F^{a^{\prime}_{L-1} \sigma}_{\sigma a_{L-1}} \right)_{\sigma,\sigma} \left( F^{\sigma\sigma}_{\sigma\sigma} \right)_{a_{L-1},a^{\prime}_{L-1}} \nonumber \\
&&~~~~~~~~~~~~\times
\begin{tikzpicture}[baseline,vertex/.style={anchor=base,
        circle,fill=black!25,minimum size=18pt,inner sep=2pt}]

\draw[black, thick] (1,0-0.5) -- (6,0-0.5);
\draw[black, thick] (1,0) -- (1.5,0);
\draw[black, thick] (1.5,1-0.5) -- (1.5,0);
\draw[orange, thick] (1.5,0) -- (2.5,0);
\draw[black, thick] (2.5,0) -- (3.5,0);
\draw[black, thick] (2.5,0.5-0.5) -- (2.5,1-0.5);
\draw[black, thick] (3.5,1-0.5) -- (3.5,0.5-0.5);
\draw[orange, thick] (3.5,0.5-0.5) -- (4,0.5-0.5);
\draw[dotted, gray, thick] (4,0) -- (5,0);
\draw[black, thick] (5,0) -- (5.5,0);
\draw[black, thick] (5.5,1-0.5) -- (5.5,0.5-0.5);
\draw[orange, thick] (5.5,0-0.5) -- (5.5,0.5-0.5);
\draw[black, thick] (5.5,0-0.5) -- (6,0-0.5);

\draw (1,0-0.5) node[anchor = east]{$\sigma(z_{1})$};
\draw (1,0) node[anchor = east]{$\sigma(z_{2})$};
\draw (2,0) node[anchor = south]{$a^{\prime}_{1}$};
\draw (1.5,1-0.5) node[anchor = south]{$\sigma(z_{3})$};
\draw (3,0) node[anchor = south]{$\sigma$};

\draw (2.5,1-0.5) node[anchor = south]{$\sigma(z_{4})$};
\draw (4,0) node[anchor = south]{$a^{\prime}_{2}$};
\draw (3.5,1-0.5) node[anchor = south]{$\sigma(z_{5})$};
\draw (5,0) node[anchor = south]{$\sigma$};

\draw (5.5,0-0.25) node[anchor = east]{$a^{\prime}_{L-1}$};
\draw (5.5,1-0.5) node[anchor = south]{$\sigma(z_{N-1})$};
\draw (6,0-0.5) node[anchor = west]{$\sigma(z_{N})$};

\end{tikzpicture} \nonumber \\
&\equiv& \sum_{a^{\prime}_{1} \ldots a^{\prime}_{L-1}} \left( \widetilde{\mathsf{D}}_{(L-1)} \right)_{a_{1} \ldots a_{L-1}, a^{\prime}_{1} \ldots a^{\prime}_{L-1}} \mathcal{F}_{\mathbf{p}^{\prime}}(z_{2}, \ldots, z_{N}, z_{1}),
\end{eqnarray}
where
\begin{equation}
\label{eq:definition-generalised-translation-Z2}
    \left( \widetilde{\mathsf{D}}_{(L-1)} \right)_{a_{1} \ldots a_{L-1}, a^{\prime}_{1} \ldots a^{\prime}_{L-1}} = \left( F^{\sigma\sigma}_{\sigma\sigma} \right)_{a_{1},a^{\prime}_{1}} \left( F^{a^{\prime}_{1} \sigma}_{\sigma a_{2}} \right)_{\sigma,\sigma} \left( F^{\sigma\sigma}_{\sigma\sigma} \right)_{a_{2},a^{\prime}_{2}} \cdots \left( F^{a^{\prime}_{L-1} \sigma}_{\sigma a_{L-1}} \right)_{\sigma,\sigma} \left( F^{\sigma\sigma}_{\sigma\sigma} \right)_{a_{L-1},a^{\prime}_{L-1}}
\end{equation}
are the matrix elements of the operator\footnote{Here, the subscript $(L-1)$ of $\widetilde{\mathsf{D}}$ indicates that this operator acts on the $(L-1)$ qubits with indices $l = 1, \ldots, L-1$. The same applies to the operators $\mathsf{R}$ and $\eta$ below.} $\widetilde{\mathsf{D}}_{(L-1)}$ in the basis specified above, and
\begin{equation}
\label{eq:conformal-block-transformed}
    \mathcal{F}_{\mathbf{p}^{\prime}}(z_{2}, \ldots, z_{N}, z_{1}) = 
\begin{tikzpicture}[baseline,vertex/.style={anchor=base,circle,fill=black!25,minimum size=18pt,inner sep=2pt}]

\draw[black, thick] (0,0-0.5) -- (0.5,0-0.5);
\draw[black, thick] (0.5,1-0.5) -- (0.5,0-0.5);
\draw[black, thick] (0.5,0-0.5) -- (1.5,0-0.5);
\draw[black, thick] (1.5,1-0.5) -- (1.5,0-0.5);
\draw[orange, thick] (1.5,0-0.5) -- (2.5,0-0.5);
\draw[black, thick] (2.5,0-0.5) -- (2.5,1-0.5);
\draw[black, thick] (2.5,0-0.5) -- (3.5,0-0.5);
\draw[black, thick] (3.5,1-0.5) -- (3.5,0-0.5);
\draw[orange, thick] (3.5,0-0.5) -- (4,0-0.5);
\draw[dotted, gray, thick] (4,0-0.5) -- (5,0-0.5);
\draw[orange, thick] (5,0-0.5) -- (5.5,0-0.5);
\draw[black, thick] (5.5,1-0.5) -- (5.5,0-0.5);
\draw[black, thick] (5.5,0-0.5) -- (6.5,0-0.5);
\draw[black, thick] (6.5,0-0.5) -- (6.5,1-0.5);
\draw[black, thick] (6.5,0-0.5) -- (7,0-0.5);

\draw (0,0-0.5) node[anchor = north]{$\boldsymbol{1}$};
\draw (0.5,1-0.5) node[anchor = south]{$\sigma(z_{2})$};
\draw (1,0-0.5) node[anchor = north]{$\sigma$};
\draw (1.5,1-0.5) node[anchor = south]{$\sigma(z_{3})$};
\draw (2,0-0.5) node[anchor = north]{$a^{\prime}_{1}$};

\draw (2.5,1-0.5) node[anchor = south]{$\sigma(z_{4})$};
\draw (3,0-0.5) node[anchor = north]{$\sigma$};
\draw (3.5,1-0.5) node[anchor = south]{$\sigma(z_{5})$};
\draw (4,0-0.5) node[anchor = north]{$a^{\prime}_{2}$};

\draw (5.5,0-0.5) node[anchor = north]{$a^{\prime}_{L-1}$};
\draw (5.5,1-0.5) node[anchor = south]{$\sigma(z_{N})$};
\draw (6,0-0.5) node[anchor = north]{$\quad\sigma$};
\draw (6.5,1-0.5) node[anchor = south]{$\sigma(z_{1})$};
\draw (7,0-0.5) node[anchor = north]{$\boldsymbol{1}$};

\end{tikzpicture},
\end{equation}
in which $\mathbf{p}^{\prime} \equiv (p_{1}^{\prime}, \ldots, p_{L-1}^{\prime})$ with the identification $a^{\prime}_{l} = \boldsymbol{1}$ (resp. $a^{\prime}_{l} = \chi$) for $p^{\prime}_{l} = 0$ (resp. $p^{\prime}_{l} = 1$). We observe that the coordinates in $\mathcal{F}_{\mathbf{p}}(z_{1}, \ldots, z_{N})$ get permuted to $(z_{2}, \ldots, z_{N}, z_{1})$ in~\eqref{eq:conformal-block-transformed}; as $N=2L$, the transformation implemented by $\widetilde{\mathsf{D}}_{(L-1)}$ can be viewed as `translating by half a site'. Thus, these conformal blocks form an eigenstate of $\widetilde{\mathsf{D}}_{(L-1)}$ with eigenvalue unity when $z_{1}, \ldots, z_{N}$ are uniformly distributed on a circle.

From~\eqref{eq:F-symbols-Z2}, it is easy to see that $\widetilde{\mathsf{D}}_{(L-1)}$ is a unitary operator. Indeed, in terms of the Hadamard
\begin{equation}
    \mathsf{H}_{l} = \frac{1}{\sqrt{2}} \left( X_{l} + Z_{l} \right) = I \otimes \cdots \otimes I \otimes \underset{\text{($l$-th qubit)}}{F^{\sigma\sigma}_{\sigma\sigma}} \otimes I \otimes \cdots \otimes I
\end{equation}
and controlled Z
\begin{equation}
    \mathsf{CZ}_{l,l+1} = \frac{1}{2} \left( 1 + Z_{l} + Z_{l+1} - Z_{l}Z_{l+1} \right)
\end{equation}
quantum gates, where $I$ is the $2 \times 2$ identity matrix, one readily finds that $\widetilde{\mathsf{D}}_{(L-1)}$ can be expressed as a Clifford circuit:
\begin{equation}
\label{eq:unitary-movement}
    \widetilde{\mathsf{D}}_{(L-1)} = \mathsf{H}_{1} \mathsf{CZ}_{1,2} \mathsf{H}_{2} \mathsf{CZ}_{2,3} \cdots \mathsf{H}_{L-2} \mathsf{CZ}_{L-2,L-1} \mathsf{H}_{L-1}.
\end{equation}
We shall see in the next subsection that the half-step translation operator, after projecting into the NS sector, gives the KW symmetry operator for the critical Ising chain.

\subsection{Half-step translation symmetry as Kramers-Wannier self-duality}

We would like to unravel the significance of $\widetilde{\mathsf{D}}_{(L-1)}$ by examining its action on the model state~\eqref{eq:Ising-GS-Ansatz}. For this purpose, we need the operator relating the `mutually dual' bases $\{ | p_{1}, \ldots, p_{L} \rangle \}$ and $\{ | \widehat{p}_{1}, \ldots, \widehat{p}_{L} \rangle \}$. It can be verified that the desired operator is given by
\begin{equation}
    \mathsf{R}_{(L-1)} \equiv \mathsf{CNOT}_{1,2} \mathsf{CNOT}_{2,3} \cdots \mathsf{CNOT}_{L-2,L-1} = \left( \prod_{l=1}^{L-1} \mathsf{H}_{l} \right) \widetilde{\mathsf{D}}_{(L-1)},
\end{equation}
where
\begin{equation}
    \mathsf{CNOT}_{l,l+1} = \frac{1}{2} \left( 1 + Z_{l} \right) + \frac{1}{2} \left( 1 - Z_{l} \right) X_{l+1} = \mathsf{H}_{l+1} \mathsf{CZ}_{l,l+1} \mathsf{H}_{l+1}
\end{equation}
is the controlled NOT gate. $\mathsf{R}_{(L-1)}$ is a unitary operator; however, it does not act within the NS sector (recall that $| \Psi \rangle$ belongs to the NS sector by construction,~$P_{+} | \Psi \rangle = | \Psi \rangle$), as $[\mathsf{R}_{(L-1)}, P_{+}] \neq 0$.\footnote{This indicates that the KW symmetry operator of the critical quantum Ising chain can be formulated as a unitary (and thus invertible) operator acting on the direct sum of the NS sector and the Ramond (R) sector. More generally, the `full' Hilbert space for the case of the $\mathbb{Z}_n$ Tambara-Yamagami fusion category incorporates boundary conditions twisted by the elements of the $\mathbb{Z}_n$ group. The operator that leaves the model state invariant becomes non-invertible due to the projection into the untwisted sector.} To remedy this, one observes that $\mathsf{R}_{(L-1)}$ does not affect the qubit with index $L$, which can be exploited as an `auxiliary qubit'; we act with the operator $P_{+}(1+X_{L})$ to ensure that each component of the state remains in the NS sector after the whole procedure. Thus, we conclude that the model state $| \Psi \rangle$ is invariant under acting with the operator
\begin{equation}
\label{eq:KW-symmetry-operator-dual-unsimplified}
    \widehat{\mathsf{D}} = P_{+} \left( 1+X_{L} \right) \mathsf{R}_{(L-1)} P_{+} \left( 1+X_{L} \right) \widetilde{\mathsf{D}}_{(L-1)} \mathsf{R}_{(L-1)}^{-1} P_{+}.
\end{equation}
The form of $\widehat{\mathsf{D}}$ in~\eqref{eq:KW-symmetry-operator-dual-unsimplified} can be simplified. We delegate the details of this simplification to appendix~\ref{append:A} (specifically, at the end of subsection~\ref{subsec:A-Z2}) and present here only the result:
\begin{equation}
\label{eq:KW-symmetry-operator-dual-simplified}
    \widehat{\mathsf{D}} = \left( \prod_{l=1}^{L} \mathsf{H}_{l} \right) \mathsf{D} \left( \prod_{l=1}^{L} \mathsf{H}_{l} \right),
\end{equation}
where
\begin{equation}
\label{eq:KW-symmetry-operator}
    \mathsf{D} = \sqrt{2}~\frac{1 + \eta_{(L)}}{2}~\widetilde{\mathsf{D}}_{(L)}~\frac{1 + \eta_{(L)}}{2}
\end{equation}
with
\begin{equation}
    \eta_{(L)} \equiv \prod_{l=1}^{L} X_{l}.
\end{equation}

We claim that $\mathsf{D}$ (or $\widehat{\mathsf{D}}$) is the symmetry operator corresponding to the KW self-duality of the critical Ising chain. Indeed, the action of $\mathsf{D}$ on the basis of $\mathbb{Z}_{2}$-invariant local operators (c.f. subsection~\ref{subsec:A-Z2} of appendix~\ref{append:A}) is as expected, and one finds $[H, \mathsf{D}] = 0$ with
\begin{equation}
    H = -\sum_{l=1}^{L-1} Z_{l}Z_{l+1} - Z_{L}Z_{1} - \sum_{l=1}^{L} X_{l} = \left( \prod_{l=1}^{L} \mathsf{H}_{l} \right) \widehat{H} \left( \prod_{l=1}^{L} \mathsf{H}_{l} \right),
\end{equation}
from which we also have $[\widehat{H}, \widehat{\mathsf{D}}] = 0$ with $\widehat{H}$ defined in~\eqref{eq:Ising-Hamiltonian}. We note that $\eta_{(L)}$ is nothing but the generator for the global $\mathbb{Z}_{2}$ symmetry of $H$.

A remark is in order. Up to a multiplicative factor of $\sqrt{2}$,~\eqref{eq:KW-symmetry-operator} is the same as one of the expressions for the KW symmetry operator given in ref.~\cite{seiberg2024b}, where it was derived by manipulating the duality defects in the critical Ising chain Hamiltonian. In ref.~\cite{seiberg2024b}, the normalisation of this operator was settled by fixing the coefficients in the composition rules (see below), whilst that in our construction was naturally fixed by the relation $\widehat{\mathsf{D}} | \Psi \rangle = | \Psi \rangle$. We have obtained this expression solely from the consideration of CFT model wave functions.

The KW symmetry operator $\mathsf{D}$ can easily be brought into the form of a matrix product operator (MPO) with periodic boundary condition. The procedure for this was presented in ref.~\cite{seiberg2024b}; to make our exposition self-contained, we also provide the details in subsection~\ref{subsec:B-Z2} of appendix~\ref{append:B}. The result is
\begin{equation}
\label{eq:KW-symmetry-operator-MPO}
    \mathsf{D} = \frac{1}{\sqrt{2}} \mathrm{Tr}_{\text{aux.}} \left( \mathbb{A}_{1} \cdots \mathbb{A}_{L} \right),
\end{equation}
where
\begin{equation}
    \mathbb{A}_{l} = \frac{1}{2} \left( \begin{array}{cc}
       (1+Z_{l})\mathsf{H}_{l}  &~ (1+Z_{l})\mathsf{H}_{l}Z_{l} \\
        (1-Z_{l})\mathsf{H}_{l} &~ (1-Z_{l})\mathsf{H}_{l}Z_{l}
    \end{array} \right)
\end{equation}
is a matrix with operator-valued entries. The periodic boundary condition is imposed by taking the trace $\mathrm{Tr}_{\text{aux.}}$ over the auxiliary (or `virtual') space (namely, the matrix indices). We see that the physical and auxiliary dimensions of this MPO are both $2$.

We have seen that $\mathsf{D}$ is proportional to the projection of the unitary operator $\widetilde{\mathsf{D}}_{(L)}$ into the NS sector. It becomes no longer unitary, as is evident from the non-invertible composition law in the `fusion algebra' involving $\mathsf{D}$. First, it is clear from~\eqref{eq:KW-symmetry-operator} that the composition of the KW symmetry operator with the global $\mathbb{Z}_2$ transformation is simply
\begin{equation}
     \eta_{(L)}\mathsf{D} = \mathsf{D}\eta_{(L)} = \mathsf{D}.
\end{equation}
The composition of $\mathsf{D}$ with itself is more interesting. This, according to the intuition of `translating by half a lattice site' we found in subsection~\ref{subsec:Ising-model-wave-function}, is naturally expected to be related to the `ordinary' translation operation. Apparently, this composition is non-invertible, as the operator $\mathsf{D}^2$ annihilates any state that is odd under the action of $\eta_{(L)}$. Using the identities given in subsection~\ref{subsec:A-Z2} of appendix~\ref{append:A}, on the other hand, one finds that $\mathsf{D}^2$ acts on the basis of $\mathbb{Z}_2$-invariant local operators as
\begin{equation}
    \mathsf{D}^{2} X_{l} = X_{l+1~(\mathrm{mod}~L)} \mathsf{D}^{2}
\end{equation}
and
\begin{equation}
    \mathsf{D}^{2} Z_{l} Z_{l+1~(\mathrm{mod}~L)} = Z_{l+1~(\mathrm{mod}~L)} Z_{l+2~(\mathrm{mod}~L)} \mathsf{D}^{2}.
\end{equation}
Thus, we conclude that
\begin{equation}
    \mathsf{D}^{2} = \frac{1}{2}~T \left( 1 + \eta_{(L)} \right),
\end{equation}
where $T$ is the translation operator satisfying
\begin{equation}
    T | p_{1}, p_{2}, \ldots, p_{L} \rangle = | p_{L}, p_{1}, \ldots, p_{L-1} \rangle.
\end{equation}
We observe that these composition laws bear a close resemblance to the corresponding fusion rules in the $\mathbb{Z}_2$ Tambara-Yamagami category. However, a crucial difference from the case of line operators in the continuum is that the composition laws are now mixed with lattice translations.

\section{\texorpdfstring{$\mathbb{Z}_{3}$}{TEXT} Kramers-Wannier self-duality and the quantum \texorpdfstring{$3$}{TEXT}-state Potts chain}
\label{sec:3-state-Potts}

\subsection{\texorpdfstring{$3$}{TEXT}-state Potts CFT and the \texorpdfstring{$\mathbb{Z}_{3}$}{TEXT} Tambara-Yamagami category}

The CFT for the scaling limit of the critical $3$-state Potts model can be described by the Virasoro minimal model $\mathcal{M}(5,6)$ with central charge $4/5$. The latter theory has 12 primary fields; their physical significance and operator algebra was studied in refs.~\cite{dotsenko1984,fateev1987}. Here, we merely mention a main complication of this CFT compared with the Ising case: its modular invariant is~\emph{non-diagonal}~\cite{francesco1997}; some of the Virasoro irreps do not appear in the partition function, and others appear twice. This indicates that the chiral algebras can be extended, as it has been shown for unitary rational CFTs that the modular invariants can be brought into the diagonal form up to an automorphism of the fusion rules with respect to the maximally extended chiral algebra~\cite{dijkgraaf1988}. Indeed, the $3$-state Potts CFT~\emph{is} diagonal when regarded as a rational CFT with $W_{3}$ algebra symmetry, which is generated by the primaries with conformal spin $\pm 3$ and contains the Virasoro algebra. The number of primaries in the $3$-state Potts CFT with respect to the $W_{3}$ algebra is reduced to 6.

There are five~\emph{primitive} topological line operators, including the trivial one, from which all simple topological lines in the $3$-state Potts CFT can be generated by composition. We focus in the following on the simple lines $\mathcal{L}_{\boldsymbol{1}}$, $\mathcal{L}_{\eta}$, $\mathcal{L}_{\eta^{2}} = \mathcal{L}_{\eta}^{2}$ and $\mathcal{L}_{\mathcal{N}}$, which close among themselves under composition and form the $\mathbb{Z}_{3}$ Tambara-Yamagami fusion category:
\begin{equation}
\label{eq:Z3-TY-fusion-rules}
    \mathcal{L}_{\eta}^{3} = \mathcal{L}_{\boldsymbol{1}}, \quad \mathcal{L}_{\eta}\mathcal{L}_{\mathcal{N}} = \mathcal{L}_{\mathcal{N}}\mathcal{L}_{\eta} = \mathcal{L}_{\mathcal{N}}, \quad \mathcal{L}_{\mathcal{N}}^{2} = \mathcal{L}_{\boldsymbol{1}} + \mathcal{L}_{\eta} + \mathcal{L}_{\eta^{2}},
\end{equation}
for which the non-trivial $F$-symbols are~\cite{chang2019}
\begin{align}
\label{eq:F-symbols-Z3}
    &\left( F^{\mathcal{N}\eta}_{\eta\mathcal{N}} \right)_{\mathcal{N},\mathcal{N}} = \left( F_{\mathcal{N}\eta}^{\eta\mathcal{N}} \right)_{\mathcal{N},\mathcal{N}}^{*} = \left( F^{\mathcal{N}\eta^{2}}_{\eta^{2}\mathcal{N}} \right)_{\mathcal{N},\mathcal{N}} = \left( F_{\mathcal{N}\eta^{2}}^{\eta^{2}\mathcal{N}} \right)_{\mathcal{N},\mathcal{N}}^{*} \nonumber \\
    &= \left( F^{\eta\mathcal{N}}_{\mathcal{N}\eta^{2}} \right)_{\mathcal{N},\mathcal{N}} = \left( F_{\eta\mathcal{N}}^{\mathcal{N}\eta^{2}} \right)_{\mathcal{N},\mathcal{N}}^{*} = \omega, \nonumber \\
    &F^{\mathcal{N}\mathcal{N}}_{\mathcal{N}\mathcal{N}} = \frac{1}{\sqrt{3}} \left( \begin{array}{ccc}
       1  &~ 1 &~ 1 \\
       1  &~ \omega^{*} &~ \omega \\
       1  &~ \omega &~ \omega^{*}
    \end{array} \right),
\end{align}
where $\omega = \mathrm{e}^{2\pi\mathrm{i}/3}$, and the rows and columns of $F^{\mathcal{N}\mathcal{N}}_{\mathcal{N}\mathcal{N}}$ are arranged in the order $(\boldsymbol{1}, \eta, \eta^{2})$. It is worth emphasising that while $\mathcal{L}_{\boldsymbol{1}}$, $\mathcal{L}_{\eta}$ and $\mathcal{L}_{\eta^{2}}$ are (invertible) Verlinde lines of the theory with extended chiral algebras, the KW duality line $\mathcal{L}_\mathcal{N}$ is~\emph{not} a Verlinde line and does not correspond to a chiral primary, as it does not commute with the $W_{3}$ algebra. We note in passing that one of the two remaining primitive topological lines is invertible and associated with the charge conjugation; the other one, obeying a non-invertible Fibonacci fusion rule, is another Verlinde line.

\subsection{Symmetry operator and the lattice Hamiltonian}

The fact we mentioned at the end of the last subsection that $\mathcal{L}_\mathcal{N}$ is not a Verlinde line of the $3$-state Potts CFT (regarded as a diagonal theory with respect to the $W_{3}$ algebra) seems to suggest that the model wave function with the desired KW self-dual property cannot be expressed as a conformal block of certain primary CVOs. Nonetheless, defining the half-step translation symmetry does not require the model wave function to be in the form of a conformal block or its explicit expression to be known. In fact, the definition~\eqref{eq:definition-generalised-translation-Z2} in terms of the $F$-symbols in the $\mathbb{Z}_2$ Tambara-Yamagami category facilitates its generalisation to other fusion categories. To highlight the parallel between the following arguments and those in the $\mathbb{Z}_2$ case, we use in this subsection the same notations as in section~\ref{sec:Ising} as long as no confusion is incurred.

\begin{figure}[htbp]
\centering
\includegraphics[width=0.618\textwidth]{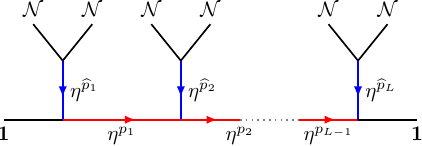}
\caption{Fusion trees in the $\mathbb{Z}_3$ Tambara-Yamagami category in which the basis $\{ | \widehat{p}_{1}, \ldots, \widehat{p}_{L} \rangle \}$ for the model wave function is encoded. Here, the height variables $p_{l}, \widehat{p}_{l} = 0, 1, 2$, satisfying $\widehat{p}_{l} = p_{l} - p_{l-1}~(\mathrm{mod}~3)~\text{with}~p_{0}  = p_{L} \equiv 0$. \label{fig:Z3-fusion-trees}}
\end{figure}

The basis for the model wave function, $\{ | \widehat{p}_{1}, \ldots, \widehat{p}_{L} \rangle \}$, is defined by the fusion trees in the $\mathbb{Z}_3$ Tambara-Yamagami category, as shown in figure~\ref{fig:Z3-fusion-trees}. Compared with figure~\ref{fig:Ising-fusion-trees}(a), the only difference is that now the height variables $p_{l}, \widehat{p}_{l} = 0, 1, 2$; as $\mathcal{L}_{\eta}$ and $\mathcal{L}_{\eta^{2}}$ are not self-conjugate, one has to choose an orientation for each of the corresponding lines. This is specified by the arrows in figure~\ref{fig:Z3-fusion-trees}. Motivated by~\eqref{eq:definition-generalised-translation-Z2}, we immediately write down the prescription for the $\mathbb{Z}_3$ version of the half-step translation operator:
\begin{align}
\label{eq:definition-generalised-translation-Z3}
    \left( \widetilde{\mathsf{D}}_{(L-1)} \right)_{a_{1} \ldots a_{L-1}, a^{\prime}_{1} \ldots a^{\prime}_{L-1}} =~&\left( F^{\mathcal{N}\mathcal{N}}_{\mathcal{N}\mathcal{N}} \right)_{a_{1},a^{\prime}_{1}} \left( F^{a^{\prime}_{1} \mathcal{N}}_{\mathcal{N} a_{2}} \right)_{\mathcal{N},\mathcal{N}} \left( F^{\mathcal{N}\mathcal{N}}_{\mathcal{N}\mathcal{N}} \right)_{a_{2},a^{\prime}_{2}} \nonumber \\
    &\times \cdots \left( F^{a^{\prime}_{L-1} \mathcal{N}}_{\mathcal{N} a_{L-1}} \right)_{\mathcal{N},\mathcal{N}} \left( F^{\mathcal{N}\mathcal{N}}_{\mathcal{N}\mathcal{N}} \right)_{a_{L-1},a^{\prime}_{L-1}},
\end{align}
in which $a_{l} \equiv \eta^{p_{l}}$. By inspecting the $F$-symbols given in~\eqref{eq:F-symbols-Z3}, we find that $\widetilde{\mathsf{D}}_{(L-1)}$ can be expressed as a circuit consisting of qu\emph{tr}it gates:
\begin{equation}
\label{eq:unitary-movement-Z3}
    \widetilde{\mathsf{D}}_{(L-1)} = \mathsf{U}_{1} \mathsf{U}_{1,2} \mathsf{U}_{2} \mathsf{U}_{2,3} \cdots \mathsf{U}_{L-2} \mathsf{U}_{L-2,L-1} \mathsf{U}_{L-1},
\end{equation}
where
\begin{subequations}
\begin{equation}
    \mathsf{U}_{l} = I \otimes \cdots \otimes I \otimes \underset{\text{($l$-th qutrit)}}{F^{\mathcal{N}\mathcal{N}}_{\mathcal{N}\mathcal{N}}} \otimes I \otimes \cdots \otimes I, \quad l = 1, \ldots, L-1
\end{equation}
with $I$ the $3 \times 3$ identity matrix, and
\begin{align}
    \mathsf{U}_{l,l+1} =~&\mathrm{diag}(1, 1, 1, 1, \omega, \omega^{*}, 1, \omega^{*}, \omega)_{l,l+1} \nonumber \\ 
    =~&\frac{1}{3} \Big( 1 + \sigma_{l} + \sigma_{l}^{\dagger} + \sigma_{l+1} + \sigma_{l+1}^{\dagger} \nonumber \\
    &+ \omega^{*} \sigma_{l} \sigma_{l+1} + \omega \sigma_{l} \sigma_{l+1}^{\dagger} + \omega \sigma_{l}^{\dagger} \sigma_{l+1} + \omega^{*} \sigma_{l}^{\dagger} \sigma_{l+1}^{\dagger} \Big), \quad l = 1, \ldots, L-2,
\end{align}
in which
\begin{equation}
\label{eq:sigma-definition}
    \sigma_{l} = \mathrm{diag}(1, \omega, \omega^{*})_{l}.
\end{equation}
\end{subequations}

We~\emph{require} that the model state $| \Psi \rangle$ (for which the explicit form is inconsequential to us) is invariant under the action of the half-step translation operation. The two sets of height variables $\{ p_{l} \}$ and $\{ \widehat{p}_{l} \}$ are related to each other via
\begin{equation}
    \widehat{p}_{l} = p_{l} - p_{l-1}~(\mathrm{mod}~3),~l = 1, \ldots, L,~\text{where}~p_{0}  = p_{L} \equiv 0.
\end{equation}
Again, we need the operator relating the bases $\{ | p_{1}, \ldots, p_{L} \rangle \}$ and $\{ | \widehat{p}_{1}, \ldots, \widehat{p}_{L} \rangle \}$, which, as can be verified, is given by
\begin{equation}
    \mathsf{R}_{(L-1)} = \mathsf{R}_{1,2} \mathsf{R}_{2,3} \cdots \mathsf{R}_{L-2,L-1} = \left( \prod_{l=1}^{L-1} \mathsf{U}_{l} \right)^{-1} \widetilde{\mathsf{D}}_{(L-1)}
\end{equation}
with
\begin{align}
\label{eq:CNOT-qutrit}
    \mathsf{R}_{l,l+1} =&~\frac{1}{3} \Big( 1 + \sigma_{l} + \sigma_{l}^{\dagger} + \tau_{l+1} + \tau_{l+1}^{\dagger} \nonumber \\
    &+ \omega\sigma_{l}\tau_{l+1} + \omega^{*}\sigma_{l}\tau_{l+1}^{\dagger} + \omega^{*}\sigma_{l}^{\dagger}\tau_{l+1} + \omega\sigma_{l}^{\dagger}\tau_{l+1}^{\dagger} \Big) \nonumber \\
    =&~\mathsf{U}_{l+1}^{-1} \mathsf{U}_{l,l+1} \mathsf{U}_{l+1}, \quad l = 1, \ldots, L-2,
\end{align}
where
\begin{equation}
\label{eq:tau-definition}
    \tau_{l} = \mathsf{U}_{l}^{-1} \sigma^{\dagger}_{l} \mathsf{U}_{l} = \left( \begin{array}{ccc}
       0  &~ 0 &~ 1 \\
       1  &~ 0 &~ 0 \\
       0  &~ 1 &~ 0
    \end{array} \right)_{l}.
\end{equation}
The obvious constraint $\sum_{l=1}^{L} \widehat{p}_{l} = 0~(\mathrm{mod}~3)$ implies that $| \Psi \rangle$, which is a linear combination of the basis states, belongs to the sector corresponding to the projector
\begin{equation}
    P_{0} = \frac{1}{3} \left( 1 + \prod_{l=1}^{L} \sigma_{l} + \prod_{l=1}^{L} \sigma_{l}^{\dagger} \right).
\end{equation}
Since $[\mathsf{R}_{(L-1)}, P_{0}] \neq 0$, we exploit the qutrit with index $L$ as an `auxiliary qutrit' and modify the unitary operator $\widetilde{\mathsf{D}}_{(L-1)}$ to the following non-invertible one:
\begin{equation}
\label{eq:3SP-KW-symmetry-operator-dual-unsimplified}
    \widehat{\mathsf{D}} = P_{0} \left( 1 + \tau_{L} + \tau_{L}^{\dagger} \right) \mathsf{R}_{(L-1)} P_{0} \left( 1 + \tau_{L} + \tau_{L}^{\dagger} \right) \widetilde{\mathsf{D}}_{(L-1)} \mathsf{R}_{(L-1)}^{-1} P_{0},
\end{equation}
which satisfies $\widehat{\mathsf{D}} | \Psi \rangle = | \Psi \rangle$. This expression can be simplified (see subsection~\ref{subsec:A-Z3} of appendix~\ref{append:A}), yielding
\begin{equation}
\label{eq:3SP-KW-symmetry-operator-dual-simplified}
    \widehat{\mathsf{D}} = \left( \prod_{l=1}^{L} \mathsf{U}_{l} \right)^{-1} \mathsf{D} \left( \prod_{l=1}^{L} \mathsf{U}_{l} \right),
\end{equation}
in which
\begin{equation}
\label{eq:3SP-KW-symmetry-operator}
    \mathsf{D} = \sqrt{3}~\frac{1 + \eta_{(L)} + \eta_{(L)}^{\dagger}}{3} \widetilde{\mathsf{D}}_{(L)} \frac{1 + \eta_{(L)} + \eta_{(L)}^{\dagger}}{3},
\end{equation}
where now
\begin{equation}
    \eta_{(L)} \equiv \prod_{l=1}^{L} \tau_{l}
\end{equation}
generates the global $\mathbb{Z}_3$ transformation.

The action of $\mathsf{D}$ on $\mathbb{Z}_3$-invariant local operators is presented in subsection~\ref{subsec:A-Z3} of appendix~\ref{append:A}. From these relations, we can readily construct a Hermitian Hamiltonian that commutes with the operator $\mathsf{D}$:
\begin{equation}
    H = - \sum_{l=1}^{L-1} \left( \sigma_{l}^{\dagger} \sigma_{l+1} + \sigma_{l} \sigma_{l+1}^{\dagger} \right) - \left( \sigma_{L}^{\dagger} \sigma_{1} + \sigma_{L} \sigma_{1}^{\dagger} \right) - \sum_{l=1}^{L} \left( \tau_{l} + \tau_{l}^{\dagger} \right),
\end{equation}
which is precisely the critical $3$-state Potts chain. It also follows that $[\widehat{H},\widehat{D}] = 0$ with
\begin{align}
    \widehat{H} &= \left( \prod_{l=1}^{L} \mathsf{U}_{l} \right)^{-1} H \left( \prod_{l=1}^{L} \mathsf{U}_{l} \right) \nonumber \\
    &= - \sum_{l=1}^{L-1} \left( \tau_{l}^{\dagger} \tau_{l+1} + \tau_{l} \tau_{l+1}^{\dagger} \right) - \left( \tau_{L}^{\dagger} \tau_{1} + \tau_{L} \tau_{1}^{\dagger} \right) - \sum_{l=1}^{L} \left( \sigma_{l} + \sigma_{l}^{\dagger} \right).
\end{align}

As for the KW symmetry operator of the critical quantum Ising chain, that of the $3$-state Potts chain can also be written as an MPO with periodic boundary condition (see subsection~\ref{subsec:B-Z3} of appendix~\ref{append:B} for the details):
\begin{equation}
    \mathsf{D} = \frac{1}{\sqrt{3}} \mathrm{Tr}_{\text{aux.}} \left( \mathbb{A}_{1} \cdots \mathbb{A}_{L} \right),
\end{equation}
where the local tensor
\begin{equation}
    \mathbb{A}_{l} = \frac{1}{3} \left( \begin{array}{ccc}
        \left( 1 + \sigma_{l} + \sigma_{l}^{\dagger} \right) \mathsf{U}_{l} &~ \left( 1 + \sigma_{l} + \sigma_{l}^{\dagger} \right) \mathsf{U}_{l}\sigma_{l} &~ \left( 1 + \sigma_{l} + \sigma_{l}^{\dagger} \right) \mathsf{U}_{l}\sigma_{l}^{\dagger} \cr
        \\
        \left( 1 + \omega^{*}\sigma_{l} + \omega\sigma_{l}^{\dagger} \right) \mathsf{U}_{l} &~ \left( 1 + \omega^{*}\sigma_{l} + \omega\sigma_{l}^{\dagger} \right) \mathsf{U}_{l}\sigma_{l} &~ \left( 1 + \omega^{*}\sigma_{l} + \omega\sigma_{l}^{\dagger} \right) \mathsf{U}_{l}\sigma_{l}^{\dagger} \cr
        \\
        \left( 1 + \omega\sigma_{l} + \omega^{*}\sigma_{l}^{\dagger} \right) \mathsf{U}_{l} &~ \left( 1 + \omega\sigma_{l} + \omega^{*}\sigma_{l}^{\dagger} \right) \mathsf{U}_{l}\sigma_{l} &~ \left( 1 + \omega\sigma_{l} + \omega^{*}\sigma_{l}^{\dagger} \right) \mathsf{U}_{l}\sigma_{l}^{\dagger}
    \end{array}\right).
\end{equation}
The physical and auxiliary dimensions are now both $3$. Finally, the composition laws involving the KW symmetry operator are similarly derived, which read
\begin{equation}
    \eta_{(L)}^{2} = \eta_{(L)}^{\dagger}, \quad \eta_{(L)}\mathsf{D} = \mathsf{D}\eta_{(L)} = \mathsf{D}, \quad \mathsf{D}^{2} = \frac{1}{3}~T \left( 1 + \eta_{(L)} + \eta_{(L)}^{\dagger} \right)
\end{equation}
with $T$ the translation operator. They are the lattice analogue of the fusion rules~\eqref{eq:Z3-TY-fusion-rules} of the $\mathbb{Z}_3$ Tambara-Yamagami category.

\section{Relations with other approaches}
\label{sec:relationship}

In this section, we would like to compare our approach with others to the KW duality on the lattice, especially in connection with the picture of `half-step translation'. We note first that, to implement the latter, some authors introduce a dual Hilbert space in which the degrees of freedom live on the links between the sites~\cite{aasen2016,aasen2020,li2023,cao2023,sinha2024}; the KW operator maps from the original Hilbert space to its dual. Particularly, in ref.~\cite{sinha2024}, where a systematic study of the line operators in the $3$-state Potts model was carried out, the KW duality defect was constructed via the half-space gauging method~\cite{shao2024} and via techniques from integrability with the model formulated in terms of the Temperley-Lieb (TL) algebra~\cite{belletête2020}. Unlike these works, the KW symmetry operator we constructed acts in the original Hilbert space, as in refs.~\cite{seiberg2024a,seiberg2024b}. This approach has the advantage that the composition laws involving the symmetry operator can be studied within a single tensor-product Hilbert space; the price one needs to pay, however, is that the fusion algebra now mixes with lattice translations.

Through representations of the TL algebra, the half-step translation symmetry can be exploited on the level of Hamiltonians~\cite{levy1991,chulliparambil2023}. For quantum chains with periodic boundary conditions, the relevant algebra is known to be the affine TL algebra~\cite{belletête2023}, which is the periodic TL algebra, with generators $\{ e_{1}, \ldots, e_{N} \}$ satisfying
\begin{equation}
\label{eq:TL-algebra}
    e_{j}^{2} = (q + q^{-1}) e_{j}, \quad e_{j} e_{j \pm 1~(\mathrm{mod}~N)} e_{j} = e_{j}, \quad e_{j} e_{j^{\prime}} = e_{j^{\prime}} e_{j},~\vert j - j^{\prime}~(\mathrm{mod}~N) \vert \geq 2,
\end{equation}
extended by two extra generators called the shift operators, $u$ and $u^{-1}$, satisfying $ue_{j} = e_{j+1}u$ and $u^{2}e_{N-1} = e_{1}e_{2} \cdots e_{N-1}$. $q$ is a complex parameter. It turns out that the Hamiltonians of both the critical quantum Ising chain and $3$-state Potts chain can be expressed, up to a constant shift, as
\begin{equation}
    H = - \sum_{j=1}^{N} e_{j};
\end{equation}
for the Ising chain,
\begin{equation}
    e_{j} = \begin{cases}
\begin{array}{c}
\frac{1}{\sqrt{2}} \left( 1 + X_{l} \right), \\
\frac{1}{\sqrt{2}} \left( 1 + Z_{l} Z_{l+1~(\mathrm{mod}~L)} \right),
\end{array} & \begin{array}{c}
j = 2l - 1 \\
j = 2l
\end{array} \end{cases}
\end{equation}
with $q = \mathrm{e}^{\mathrm{i}\pi/4}$, while for the $3$-state Potts chain,
\begin{equation}
    e_{j} = \begin{cases}
\begin{array}{c}
\frac{1}{\sqrt{3}} \left( 1 + \tau_{l} + \tau_{l}^{\dagger} \right), \\
\frac{1}{\sqrt{3}} \left( 1 + \sigma_{l}^{\dagger} \sigma_{l+1~(\mathrm{mod}~L)} + \sigma_{l} \sigma_{l+1~(\mathrm{mod}~L)}^{\dagger} \right),
\end{array} & \begin{array}{c}
j = 2l - 1 \\
j = 2l
\end{array} \end{cases}
\end{equation}
with $q = \mathrm{e}^{\mathrm{i}\pi/6}$. As each physical site (labelled by $l = 1, \ldots, L$) is split into two `TL sites' [labelled by $j = 1, \ldots, N(=2L)$], the Hilbert space of the model in the latter representation typically no longer admits the tensor product structure. For the Ising chain, it turned out that the non-invertible operator implementing the translation by one TL site can still be expressed in terms of $\{ e_{1}, \ldots, e_{N-1} \}$, which (up to an arbitrary phase factor) reads~\cite{chulliparambil2023,seiberg2024a}
\begin{subequations}
\label{eq:KW-symmetry-operator-TL}
    \begin{equation}
        \mathsf{D}^{\prime} = \frac{1+\eta}{2} \exp{\left( \frac{\mathrm{i}\pi}{2\sqrt{2}}e_{1} \right)} \exp{\left( \frac{\mathrm{i}\pi}{2\sqrt{2}}e_{2} \right)} \cdots \exp{\left( \frac{\mathrm{i}\pi}{2\sqrt{2}}e_{N-1} \right)},
    \end{equation}
where
    \begin{equation}
        \eta = \left( \sqrt{2}e_{1} - 1 \right) \left( \sqrt{2}e_{3} - 1 \right) \cdots \left( \sqrt{2}e_{N-1} - 1 \right).
    \end{equation}
\end{subequations}
$\mathsf{D}^{\prime}$ does not involve the shift operators and acts in the original tensor-product Hilbert space. It is straightforward to verify that $\mathsf{D}^{\prime} e_{j} = e_{j+1~(\mathrm{mod}~N)} \mathsf{D}^{\prime}$, which is identical to the action of $\mathsf{D}$ [eqs.~\eqref{eq:intertwining3} and~\eqref{eq:intertwining4} in appendix~\ref{append:A}]; in particular, the projection to the NS sector is necessary to guarantee that $e_{N-1} \mapsto e_{N}$ and $e_{N} \mapsto e_{1}$. However, the translational invariance of the expression~\eqref{eq:KW-symmetry-operator-TL} is not manifest, in contrast to $\mathsf{D}$ which has been rewritten as an MPO with periodic boundary condition [eq.~\eqref{eq:KW-symmetry-operator-MPO}]. We remark that the same operation can also be realised by a product of braid operators, either in the TL~\cite{montes2017a} or Majorana fermion~\cite{chen2022} representation; this is not surprising, as the generators of the Hecke algebra, which are intimately related to those of the TL algebra, satisfy the defining relations for the braid operators~\cite{gomez1996}. It is anticipated that there exists a $\mathbb{Z}_3$ parafermion version of this construction for the $3$-state Potts chain~\cite{tu2024}. We conjecture that the equivalence between this formulation and ours results from the relation between the $B$- and $F$-symbols established by the Moore-Seiberg polynomial equations in a modular tensor category. The process of `splitting one site into two' is not involved in the approach we followed in the present work, as $\{ z_{1}, \ldots, z_{N} \}$ are simply parameters of the model wave function (c.f. section~\ref{sec:Ising}).

In ref.~\cite{seiberg2024b}, the KW symmetry operator~\eqref{eq:KW-symmetry-operator} for the critical Ising chain was derived by exploiting a general property of symmetry defects in quantum chains, namely translating around the periodic chain in the presence of a defect produces another defect line of the same type along the spatial direction, which is, of course, the symmetry operator. Specifically, acting the unitary operator $\mathsf{H}_{l}\mathsf{CZ}_{l,l+1}$ in~\eqref{eq:unitary-movement} on a KW duality defect between sites $l$ and $l+1$ translates it by one site~\cite{schutz1993,grimm1993}. The origin of these unitary gates is clear in our approach: they are nothing but the $F$-symbols in the Ising fusion category. In this respect, our approach may be regarded as viewing the above picture from the perspective of model wave functions, which does not entail introducing the lattice defects (and, in fact, even the Hamiltonian itself). We note that MPO symmetries/intertwiners constructed from $F$-symbols were also proposed in refs.~\cite{lootens2023,lootens2024}, where they arose from solutions to intertwining conditions reminiscent of the Yang-Baxter equations.

\section{Summary and outlook}
\label{sec:summary}

In summary, we have proposed a Hamiltonian-independent approach to deriving the symmetry operators associated with the KW self-duality for critical quantum chains, leveraging model wave functions. The derivation is based on the minimal assumption, i.e., what we have referred to as the half-step translation symmetry, regarding the model wave functions. The KW symmetry operator is directly related to infrared data, namely the $F$-symbols, expressed as certain unitary quantum gates, in the fusion category of topological line operators in the CFT describing the scaling limit of the critical model, and admits the form of a translationally invariant MPO. Well-known critical models can subsequently be obtained as Hamiltonians that commute with the corresponding KW symmetry operators. As concrete examples, the transverse-field Ising chain and $3$-state Potts chain are considered, producing both known and new results. The (non-)invertible composition laws of the symmetry operators in these models resemble the fusion rules in the $\mathbb{Z}_2$ and $\mathbb{Z}_3$ Tambara-Yamagami categories, from which these KW symmetry operators are constructed, respectively, but with the important difference that lattice translations are now involved. The connections and differences between our approach and several other existing ones are also briefly discussed.

Let us note a few relevant open problems that warrant future exploration. Firstly, we have focused on the case where the line operators are only along the spatial direction. It is of interest to study the situations in which there are line operators along both the spatial and the temporal directions; this corresponds to the symmetry operators of quantum chains with defect-twisted boundary conditions. In our formalism, this becomes the question of how the twisted boundary conditions manifest themselves in model wave functions. For the latter constructed from conformal blocks, in particular, we expect that this can be achieved by inserting certain boundary-condition-changing fields into the correlators.

The relationship between our approach and others also deserves further investigation. As we have mentioned in section~\ref{sec:relationship}, one example is the $\mathbb{Z}_3$ parafermion formulation of the KW symmetry operator for the critical quantum $3$-state Potts chain, which will complement the present work.

Another natural direction to pursue is to consider non-invertible symmetries described by more general fusion categories. In fact, given a generic fusion category, the `half-step translation' operator [generalising~\eqref{eq:definition-generalised-translation-Z2}] can be formally defined for a fusion tree in the multiperipheral basis. However, its physical meaning (e.g. the action on local operators) is~\emph{a priori} not clear. Moreover, it remains an open question which fusion categories give rise to lattice non-invertible symmetries --- potentially involving translations --- on a~\emph{tensor-product} Hilbert space. For the Tambara-Yamagami category $\mathrm{TY}(G,\chi,\epsilon)$ modelling the invariance under gauging $G$ (an Abelian finite group, which is a non-anomalous global symmetry of the CFT under consideration) with the so-called bicharacter $\chi$ and the Frobenius-Schur indicator $\epsilon$~\cite{tambara1998}, the situation is simpler due to the fusion rules. The KW self-duality examined in this work corresponds to the special cases $G = \mathbb{Z}_{2}, \mathbb{Z}_{3}$ and $\epsilon = +1$ (there is a unique, trivial, $\chi$ for the group $\mathbb{Z}_n$~\cite{chang2019}; the choice of $\epsilon$ is immaterial for the~\emph{lattice} symmetry operators~\cite{seiberg2024b}, see also footnote~\ref{ftnt:FS}). As illustrated in figure~\ref{fig:Ising-fusion-trees}(b) for the example with $G = \mathbb{Z}_{2}$, the variables $p_{l},~l = 1, \ldots, L-1$ are independent of each other, in accordance with the requirement that the degrees of freedom living on different sites should not be subject to non-local~\emph{kinematic} constraint that cannot be expressed as a product form. We leave the investigation of the more general conditions for future works.

Finally, we hope to draw attention to a point mentioned in section~\ref{sec:Ising}, which is that the model state~\eqref{eq:Ising-GS-Ansatz} (when $z_{1}, \ldots, z_{N}$ are uniformly distributed on a circle) constructed from conformal blocks in the Ising CFT turns out to be the exact ground state of the critical quantum Ising chain. Proved in ref.~\cite{montes2017b} using the free fermion representation of the model, this is a non-trivial fact and appears to indicate some peculiarity about the Ising model. It is unclear whether this peculiarity of the Ising model is just that it admits the free fermion representation or if other mathematical structures, possibly also existing in other models, guarantee this property. To shed light on these questions, we plan to prove the same property in a more `axiomatic' way. To this end, the KW self-duality itself is not sufficient, and more powerful constraints should be exploited. For lattice models associated with the Wess-Zumino-Witten CFTs~\cite{cirac2010,nielsen2011}, the appropriate constraint was given by the decoupling of null fields (or, the Knizhnik-Zamolodchikov equations~\cite{knizhnik1984}); these do not generalise straightforwardly to the case of Virasoro minimal models, though the Coulomb gas formulation~\cite{dotsenko1984a,dotsenko1985} may provide useful insights.

\acknowledgments
We would like to thank Hong-Hao Tu and Javier Molina-Vilaplana for their valuable discussions. We also acknowledge support from the Spanish MINECO grant PID2021-127726NB-I00, the CSIC Research Platform on Quantum Technologies PTI-001, and the QUANTUM ENIA project Quantum Spain through the RTRP-Next Generation program under the framework of the Digital Spain 2026 Agenda.

\appendix
\section{4-point conformal blocks in the Ising CFT and ground state of the lattice Hamiltonian}
\label{append:O}

In this appendix, we explicitly compare the ground-state wave function of~\eqref{eq:Ising-Hamiltonian} with $2$ sites,
\begin{equation}
\label{eq:Ising-Hamiltonian-2-sites}
    \widehat{H}_{L=2} = -2X_{1}X_{2} - Z_{1} - Z_{2},
\end{equation}
to the $(N=2L=4)$-point conformal block of the chiral primary $\sigma$ in the Ising CFT.

In the basis $\{ | \widehat{p}_{1}, \widehat{p}_{2} \rangle \}$, the states in the NS sector are expressed as
\begin{equation}
    |\psi\rangle = \psi_{00} |00\rangle + \psi_{11} |11\rangle.
\end{equation}
The eigenvalue equation is obtained by acting $\widehat{H}_{L=2}$ on $|\psi\rangle$ as
\begin{equation}
    -2 \left( \begin{array}{cc}
       1  &~ 1 \\
        1 &~ -1
    \end{array} \right) \left( \begin{array}{c}
       \psi_{00} \\
        \psi_{11}
    \end{array} \right) = E \left( \begin{array}{c}
       \psi_{00} \\
        \psi_{11}
    \end{array} \right),
\end{equation}
where the energy eigenvalues are easily solved to be $E = \pm E_{\text{GS}}$ with the ground-state energy $E_{\text{GS}} = -2\sqrt{2}$, for which the wave-function coefficients satisfy $(\psi_{\text{GS}})_{11} / (\psi_{\text{GS}})_{00} = \sqrt{2} - 1$. In terms of the Hadamard gate
\begin{equation}
    \mathsf{H} = \frac{1}{\sqrt{2}} \left( \begin{array}{cc}
       1  &~ 1 \\
        1 &~ -1
    \end{array} \right),
\end{equation}
the eigenvalue equation satisfied by the ground state is simply
\begin{equation}
\label{eq:2-site-eigenvalue-equation}
    \mathsf{H} | \psi_{\text{GS}} \rangle  = | \psi_{\text{GS}} \rangle.
\end{equation}
We observe that~\eqref{eq:2-site-eigenvalue-equation} is concurrently also the condition for $| \psi_{\text{GS}} \rangle$ to be invariant under the half-step translation [c.f. eq.~\eqref{eq:definition-generalised-translation-Z2}], as $\widetilde{\mathsf{D}}_{(1)} = \mathsf{H}$ is identical to the projection of the Hamiltonian~\eqref{eq:Ising-Hamiltonian-2-sites} in the NS sector.

The conformal blocks for an arbitrary number of chiral primaries in the Ising CFT were computed in ref.~\cite{ardonne2010}, where the result for $4$-point blocks of $\sigma$ is given by
\begin{equation}
    \mathcal{F}_{p_{1}}(z_{1}, z_{2}, z_{3}, z_{4}) = \left[ \sqrt{\left( z_{1} - z_{3} \right) \left( z_{2} - z_{4} \right)} + (-1)^{p_{1}}\sqrt{\left( z_{1} - z_{4} \right) \left( z_{2} - z_{3} \right)} \right]^{1/2}
\end{equation}
up to an unimportant multiplicative constant, where $p_{1} = 0, 1$ is the only internal fusion channel [c.f. eq.~\eqref{eq:Ising-CB}]. When the $z_{j}$, $j = 1, \ldots, 4$, are uniformly distributed on the unit circle, we choose $z_{j} = \mathrm{e}^{\mathrm{i}\pi(j-1)/2}$ and find
\begin{equation}
    \frac{\mathcal{F}_{1}(z_{1}, z_{2}, z_{3}, z_{4})}{\mathcal{F}_{0}(z_{1}, z_{2}, z_{3}, z_{4})} = \sqrt{2} - 1 = \frac{\left( \psi_{\text{GS}}\right)_{11}}{\left( \psi_{\text{GS}}\right)_{00}}.
\end{equation}
Thus, we have verified for the case $L=2$ that the ground state of the critical Ising chain, $| \psi_{\text{GS}} \rangle$, agrees with the model state $| \Psi \rangle$ defined in~\eqref{eq:Ising-GS-Ansatz}.

\section{Kramers-Wannier duality transformation on local operators}
\label{append:A}

In this appendix, we derive the action of $\mathbb{Z}_2$ and $\mathbb{Z}_3$ KW duality transformations on the corresponding basis of symmetric local operators.

\subsection{\texorpdfstring{$\mathbb{Z}_{2}$}{TEXT} case}
\label{subsec:A-Z2}

It is straightforward to verify that
\begin{equation}
    \widetilde{\mathsf{D}}_{(L)} X_{l} = \begin{cases}
\begin{array}{c}
Z_{l}Z_{l+1} \widetilde{\mathsf{D}}_{(L)}, \\
Z_{L} \widetilde{\mathsf{D}}_{(L)},
\end{array} & \begin{array}{c}
l = 1, \ldots, L-1\\
l = L
\end{array} \end{cases}
\end{equation}
and
\begin{equation}
\label{eq:intertwining2}
    \widetilde{\mathsf{D}}_{(L)} Z_{l} = \left( \prod_{m=1}^{l} X_{m} \right) \widetilde{\mathsf{D}}_{(L)}, \quad l = 1, \ldots, L,
\end{equation}
from which it follows
\begin{subequations}
\begin{equation}
    \widetilde{\mathsf{D}}_{(L)} Z_{l} Z_{l+1} = X_{l+1} \widetilde{\mathsf{D}}_{(L)}, \quad l = 1, \ldots, L-1,
\end{equation}
\begin{equation}
    \widetilde{\mathsf{D}}_{(L)} Z_{L} Z_{1} = \left(\prod_{l=2}^{L} X_{l} \right) \widetilde{\mathsf{D}}_{(L)}.
\end{equation}
\end{subequations}
Finally, using~\eqref{eq:KW-symmetry-operator},
one finds that
\begin{equation}
\label{eq:intertwining3}
    \mathsf{D} X_{l} = Z_{l} Z_{l+1~(\mathrm{mod}~L)} \mathsf{D}
\end{equation}
and
\begin{equation}
\label{eq:intertwining4}
    \mathsf{D} Z_{l} Z_{l+1~(\mathrm{mod}~L)} = X_{l+1~(\mathrm{mod}~L)} \mathsf{D}.
\end{equation}
As an example, we illustrate explicitly the calculation of $\mathsf{D} X_{L}$ as follows:
\begin{align}
    \mathsf{D} X_{L} &= \sqrt{2}~\frac{1+\eta_{(L)}}{2}~\widetilde{\mathsf{D}}_{(L)}~\frac{1+\eta_{(L)}}{2} X_{L} = \sqrt{2}~\frac{1+\eta_{(L)}}{2} Z_{L} \widetilde{\mathsf{D}}_{(L)}~\frac{1+\eta_{(L)}}{2} \nonumber \\
    &= \sqrt{2} Z_{L} \frac{1-\eta_{(L)}}{2}~\widetilde{\mathsf{D}}_{(L)}~\frac{1+\eta_{(L)}}{2} \nonumber \\
    &= \sqrt{2} Z_{L} Z_{1} \frac{1+\eta_{(L)}}{2} Z_{1} \widetilde{\mathsf{D}}_{(L)}~\frac{1+\eta_{(L)}}{2} \nonumber \\
    &= \sqrt{2} Z_{L} Z_{1} \frac{1+\eta_{(L)}}{2}~\widetilde{\mathsf{D}}_{(L)} \eta_{(L)}~\frac{1+\eta_{(L)}}{2} \nonumber \\
    &= Z_{L} Z_{1} \mathsf{D},
\end{align}
where we have used $Z_{1} \widetilde{\mathsf{D}}_{(L)} = \widetilde{\mathsf{D}}_{(L)} \eta_{(L)}$.

We now display the derivation of the expression~\eqref{eq:KW-symmetry-operator-dual-simplified} for $\widehat{\mathsf{D}}$, using the identities listed above. Substituting
\begin{equation}
    \mathsf{R}_{(L-1)} = \left( \prod_{l=1}^{L-1} \mathsf{H}_{l} \right) \widetilde{\mathsf{D}}_{(L-1)}
\end{equation}
into~\eqref{eq:KW-symmetry-operator-dual-unsimplified}, we obtain
\begin{align}
\label{eq:KW-symmetry-operator-dual-simplification}
    \widehat{\mathsf{D}} &= P_{+} \left( 1+X_{L} \right) \left( \prod_{l=1}^{L-1} \mathsf{H}_{l} \right) \widetilde{\mathsf{D}}_{(L-1)} P_{+} \left( 1+X_{L} \right) \left( \prod_{l=1}^{L-1} \mathsf{H}_{l} \right) P_{+} \nonumber \\
    &= \left( \prod_{l=1}^{L-1} \mathsf{H}_{l} \right) \frac{1 + \eta_{(L-1)}Z_{L}}{2} \widetilde{\mathsf{D}}_{(L-1)} \left( 1+X_{L} \right) P_{+} \left( 1+X_{L} \right) \frac{1 + \eta_{(L-1)}Z_{L}}{2} \left( \prod_{l=1}^{L-1} \mathsf{H}_{l} \right) \nonumber \\
    &= \left( \prod_{l=1}^{L-1} \mathsf{H}_{l} \right) \frac{1 + \eta_{(L-1)}Z_{L}}{2} \widetilde{\mathsf{D}}_{(L-1)} \left( 1+X_{L} \right) \frac{1 + \eta_{(L-1)}Z_{L}}{2} \left( \prod_{l=1}^{L-1} \mathsf{H}_{l} \right) \nonumber \\
    &= \left( \prod_{l=1}^{L} \mathsf{H}_{l} \right) \frac{1 + \eta_{(L)}}{2} \widetilde{\mathsf{D}}_{(L-1)} \left( 1+Z_{L} \right) \frac{1 + \eta_{(L)}}{2} \left( \prod_{l=1}^{L} \mathsf{H}_{l} \right) \nonumber \\
    &= \left( \prod_{l=1}^{L} \mathsf{H}_{l} \right) \widetilde{\mathsf{D}}_{(L)} \frac{1+Z_{L}}{2} \mathsf{H}_{L} \mathsf{CZ}_{L-1,L} \left( 1+Z_{L} \right) \frac{1 + \eta_{(L)}}{2} \left( \prod_{l=1}^{L} \mathsf{H}_{l} \right) \nonumber \\
    &= \sqrt{2} \left( \prod_{l=1}^{L} \mathsf{H}_{l} \right) \widetilde{\mathsf{D}}_{(L)} \frac{1+Z_{L}}{2} \frac{1 + \eta_{(L)}}{2} \left( \prod_{l=1}^{L} \mathsf{H}_{l} \right) \nonumber \\
    &= \sqrt{2} \left( \prod_{l=1}^{L} \mathsf{H}_{l} \right) \frac{1 + \eta_{(L)}}{2} \widetilde{\mathsf{D}}_{(L)} \frac{1 + \eta_{(L)}}{2} \left( \prod_{l=1}^{L} \mathsf{H}_{l} \right) \nonumber \\
    &= \left( \prod_{l=1}^{L} \mathsf{H}_{l} \right) \mathsf{D} \left( \prod_{l=1}^{L} \mathsf{H}_{l} \right).
\end{align}
In the 4th equality in~\eqref{eq:KW-symmetry-operator-dual-simplification}, we replaced $X_{L}$ by $\mathsf{H}_{L} Z_{L} \mathsf{H}_{L}$ and `pulled out' this pair of Hadamard gates to the ends of the expression; in the 5th equality, we replaced $\widetilde{\mathsf{D}}_{(L-1)}$ by $\widetilde{\mathsf{D}}_{(L)} \mathsf{H}_{L} \mathsf{CZ}_{L-1,L}$ and used~\eqref{eq:intertwining2}; in the 6th equality, we computed explicitly the expression for $(1+Z_{L})\mathsf{H}_{L}\mathsf{CZ}_{L-1,L}(1+Z_{L})$; the 7th equality used~\eqref{eq:intertwining2} again.

\subsection{\texorpdfstring{$\mathbb{Z}_{3}$}{TEXT} case}
\label{subsec:A-Z3}

The basic on-site algebra satisfied by the operators $\sigma_{l}$ and $\tau_{l}$ [the matrix representation we use in this work is given in~\eqref{eq:sigma-definition} and~\eqref{eq:tau-definition}] is
\begin{equation}
    \sigma_{l}^{3} = \tau_{l}^{3} = 1, \quad \sigma_{l}^{2} = \sigma_{l}^{\dagger}, \quad \tau_{l}^{2} = \tau_{l}^{\dagger}, \quad \sigma_{l} \tau_{l} = \omega \tau_{l} \sigma_{l}~\text{with}~\omega = \mathrm{e}^{2\pi\mathrm{i}/3}.
\end{equation}
Utilising the basic algebra, the action of the unitaries $\mathsf{U}_{l}$ and $\mathsf{U}_{l,l+1}$ on these local operators can be derived:
\begin{align}
    &\mathsf{U}_{l} \sigma_{l} \mathsf{U}_{l}^{-1} = \tau_{l}, \quad \mathsf{U}_{l} \tau_{l} \mathsf{U}_{l}^{-1} = \sigma_{l}^{\dagger}, \nonumber \\
    &\mathsf{U}_{l,l+1} \tau_{l} \mathsf{U}_{l,l+1}^{-1} = \tau_{l} \sigma_{l+1}, \quad \mathsf{U}_{l,l+1} \tau_{l+1} \mathsf{U}_{l,l+1}^{-1} = \sigma_{l} \tau_{l+1}.
\end{align}
From these relations, one further obtains the $\mathbb{Z}_{3}$ counterpart of the identities in subsection~\ref{subsec:A-Z2}, which we list in the following.
\begin{equation}
\label{eq:intertwining1}
    \widetilde{\mathsf{D}}_{(L)} \sigma_{l} = \left( \prod_{m=1}^{l} \tau_{m} \right) \widetilde{\mathsf{D}}_{(L)}, \quad l = 1, \ldots, L,
\end{equation}
\begin{subequations}
\begin{equation}
    \widetilde{\mathsf{D}}_{(L)} \sigma_{l}^{\dagger} \sigma_{l+1} = \tau_{l+1} \widetilde{\mathsf{D}}_{(L)}, \quad l = 1, \ldots, L-1,
\end{equation}
\begin{equation}
    \widetilde{\mathsf{D}}_{(L)} \sigma_{L}^{\dagger} \sigma_{1} = \left(\prod_{l=2}^{L} \tau_{l}^{\dagger} \right) \widetilde{\mathsf{D}}_{(L)};
\end{equation}
\end{subequations}
\begin{equation}
    \widetilde{\mathsf{D}}_{(L)} \tau_{l} = \begin{cases}
\begin{array}{c}
\sigma_{l}^{\dagger} \sigma_{l+1} \widetilde{\mathsf{D}}_{(L)}, \\
\sigma_{L}^{\dagger} \widetilde{\mathsf{D}}_{(L)},
\end{array} & \begin{array}{c}
l = 1, \ldots, L-1\\
l = L
\end{array} \end{cases}.
\end{equation}
\begin{equation}
    \mathsf{D} \sigma_{l}^{\dagger} \sigma_{l+1~(\mathrm{mod}~L)} = \tau_{l+1~(\mathrm{mod}~L)} \mathsf{D};
\end{equation}
\begin{equation}
    \mathsf{D} \tau_{l} = \sigma_{l}^{\dagger} \sigma_{l+1~(\mathrm{mod}~L)} \mathsf{D}.
\end{equation}
In the derivation of the equality $\mathsf{D} \tau_{L} = \sigma_{L}^{\dagger} \sigma_{1} \mathsf{D}$, we have used $\sigma_{1}^{\dagger} \widetilde{\mathsf{D}}_{(L)} = \widetilde{\mathsf{D}}_{(L)} \eta_{(L)}$.

As an application of the above identities, one can derive the expression~\eqref{eq:3SP-KW-symmetry-operator-dual-simplified} for $\widehat{\mathsf{D}}$. Since the derivation is completely parallel to that in~\eqref{eq:KW-symmetry-operator-dual-simplification}, we spell out the steps without explanation:
\begin{align}
    \widehat{\mathsf{D}} =&~P_{0} \left( 1 + \tau_{L} + \tau_{L}^{\dagger} \right) \left( \prod_{l=1}^{L-1} \mathsf{U}_{l} \right)^{-1} \widetilde{\mathsf{D}}_{(L-1)} P_{0} \left( 1 + \tau_{L} + \tau_{L}^{\dagger} \right) \left( \prod_{l=1}^{L-1} \mathsf{U}_{l} \right) P_{0} \nonumber \\
    =&~\left( \prod_{l=1}^{L-1} \mathsf{U}_{l} \right)^{-1} \frac{1 + \eta_{(L-1)} \sigma_{L} + \eta_{(L-1)}^{\dagger} \sigma_{L}^{\dagger}}{3} \widetilde{\mathsf{D}}_{(L-1)} \left( 1 + \tau_{L} + \tau_{L}^{\dagger} \right) P_{0} \left( 1 + \tau_{L} + \tau_{L}^{\dagger} \right) \nonumber \\
    &~\times \frac{1 + \eta_{(L-1)} \sigma_{L} + \eta_{(L-1)}^{\dagger} \sigma_{L}^{\dagger}}{3} \left( \prod_{l=1}^{L-1} \mathsf{U}_{l} \right) \nonumber \\
    =&~\left( \prod_{l=1}^{L-1} \mathsf{U}_{l} \right)^{-1} \frac{1 + \eta_{(L-1)} \sigma_{L} + \eta_{(L-1)}^{\dagger} \sigma_{L}^{\dagger}}{3} \widetilde{\mathsf{D}}_{(L-1)} \left( 1 + \tau_{L} + \tau_{L}^{\dagger} \right) \frac{1 + \eta_{(L-1)} \sigma_{L} + \eta_{(L-1)}^{\dagger} \sigma_{L}^{\dagger}}{3} \nonumber \\
    &~\times \left( \prod_{l=1}^{L-1} \mathsf{U}_{l} \right) \nonumber \\
    =&~\left( \prod_{l=1}^{L} \mathsf{U}_{l} \right)^{-1} \frac{1 + \eta_{(L)} + \eta_{(L)}^{\dagger}}{3} \widetilde{\mathsf{D}}_{(L-1)} \left( 1 + \sigma_{L} + \sigma_{L}^{\dagger} \right) \frac{1 + \eta_{(L)} + \eta_{(L)}^{\dagger}}{3} \left( \prod_{l=1}^{L} \mathsf{U}_{l} \right) \nonumber \\
    =&~\left( \prod_{l=1}^{L} \mathsf{U}_{l} \right)^{-1} \widetilde{\mathsf{D}}_{(L)} \frac{1 + \sigma_{L} + \sigma_{L}^{\dagger}}{3} \mathsf{U}_{L}^{-1} \mathsf{U}_{L-1,L}^{-1} \left( 1 + \sigma_{L} + \sigma_{L}^{\dagger} \right) \frac{1 + \eta_{(L)} + \eta_{(L)}^{\dagger}}{3} \left( \prod_{l=1}^{L} \mathsf{U}_{l} \right) \nonumber \\
    =&~\sqrt{3} \left( \prod_{l=1}^{L} \mathsf{U}_{l} \right)^{-1} \widetilde{\mathsf{D}}_{(L)} \frac{1 + \sigma_{L} + \sigma_{L}^{\dagger}}{3} \frac{1 + \eta_{(L)} + \eta_{(L)}^{\dagger}}{3} \left( \prod_{l=1}^{L} \mathsf{U}_{l} \right) \nonumber \\
    =&~\sqrt{3} \left( \prod_{l=1}^{L} \mathsf{U}_{l} \right)^{-1} \frac{1 + \eta_{(L)} + \eta_{(L)}^{\dagger}}{3} \widetilde{\mathsf{D}}_{(L)} \frac{1 + \eta_{(L)} + \eta_{(L)}^{\dagger}}{3} \left( \prod_{l=1}^{L} \mathsf{U}_{l} \right) \nonumber \\
    = &~\left( \prod_{l=1}^{L} \mathsf{U}_{l} \right)^{-1} \mathsf{D} \left( \prod_{l=1}^{L} \mathsf{U}_{l} \right).
\end{align}

\section{MPO representation of the Kramers-Wannier symmetry operator}
\label{append:B}

In this appendix, we recast the KW symmetry operators for the critical quantum Ising and $3$-state Potts chains into the form of an MPO with periodic boundary condition. The details for the $\mathbb{Z}_2$ (Ising) case can also be found in ref.~\cite{seiberg2024b}.

\subsection{\texorpdfstring{$\mathbb{Z}_{2}$}{TEXT} case}
\label{subsec:B-Z2}

First, utilising the relations $Z_{1} \widetilde{\mathsf{D}}_{(L)} = \widetilde{\mathsf{D}}_{(L)} \eta_{(L)}$ and $\eta_{(L)} \widetilde{\mathsf{D}}_{(L)} = \widetilde{\mathsf{D}}_{(L)} Z_{L}$, one has
\begin{equation}
\label{eq:KW-symmetry-operator-alternative}
    \mathsf{D} = \frac{1}{\sqrt{2}} \left( \frac{1+Z_{1}}{2} \widetilde{\mathsf{D}}_{(L)} + \frac{1-Z_{1}}{2} \widetilde{\mathsf{D}}_{(L)} Z_{L} \right).
\end{equation}
Next, we express
\begin{equation}
    \widetilde{\mathsf{D}}_{(L)} = \mathsf{H}_{1} \mathsf{CZ}_{1,2} \mathsf{H}_{2} \mathsf{CZ}_{2,3} \cdots \mathsf{H}_{L-1} \mathsf{CZ}_{L-1,L} \mathsf{H}_{L}
\end{equation}
as the product of certain column and row matrices with operator-valued entries. The observation is that
\begin{equation}
    \mathsf{H}_{l} \mathsf{CZ}_{l,l+1} = \left( \begin{array}{cc}
        \mathsf{H}_{l} &~ \mathsf{H}_{l}Z_{l}
    \end{array}\right) \left( \begin{array}{c}
         \frac{1+Z_{l+1}}{2} \cr
         \\
         \frac{1-Z_{l+1}}{2}
    \end{array} \right).
\end{equation}
Therefore,~\eqref{eq:KW-symmetry-operator-alternative} can be rewritten as
\begin{align}
    \mathsf{D} =& \frac{1}{\sqrt{2}} \mathrm{Tr}_{\text{aux.}} \Bigg[ \left( \begin{array}{c}
         \frac{1+Z_{1}}{2} \cr
         \\
         \frac{1-Z_{1}}{2}
    \end{array} \right)
    \left( \begin{array}{cc}
        \mathsf{H}_{1} &~ \mathsf{H}_{1}Z_{1}
    \end{array}\right) \left( \begin{array}{c}
         \frac{1+Z_{2}}{2} \cr
         \\
         \frac{1-Z_{2}}{2}
    \end{array} \right) \cdots \nonumber \\
    &~~~~~~~~~~~~~~\times \left( \begin{array}{cc}
        \mathsf{H}_{L-1} &~ \mathsf{H}_{L-1}Z_{L-1}
    \end{array}\right) \left( \begin{array}{c}
         \frac{1+Z_{L}}{2} \cr
         \\
         \frac{1-Z_{L}}{2}
    \end{array} \right)
    \left( \begin{array}{cc}
        \mathsf{H}_{L} &~ \mathsf{H}_{L}Z_{L}
    \end{array} \right) \Bigg] \nonumber \\
    =& \frac{1}{\sqrt{2}} \mathrm{Tr}_{\text{aux.}} \left( \mathbb{A}_{1} \cdots \mathbb{A}_{L} \right),
\end{align}
where
\begin{equation}
    \mathbb{A}_{l} = \left( \begin{array}{c}
         \frac{1+Z_{l}}{2} \cr
         \\
         \frac{1-Z_{l}}{2}
    \end{array} \right)
    \left( \begin{array}{cc}
        \mathsf{H}_{l} &~ \mathsf{H}_{l}Z_{l}
    \end{array} \right) = \frac{1}{2} \left( \begin{array}{cc}
       (1+Z_{l})\mathsf{H}_{l}  &~ (1+Z_{l})\mathsf{H}_{l}Z_{l} \\
        (1-Z_{l})\mathsf{H}_{l} &~ (1-Z_{l})\mathsf{H}_{l}Z_{l}
    \end{array} \right)
\end{equation}
is the local tensor of an MPO with periodic boundary condition, for which the physical and auxiliary dimensions are both $2$.

\subsection{\texorpdfstring{$\mathbb{Z}_{3}$}{TEXT} case}
\label{subsec:B-Z3}

Utilising the relations $\eta_{(L)} \widetilde{\mathsf{D}}_{(L)} = \widetilde{\mathsf{D}}_{(L)} \sigma_{L},~\eta_{(L)}^{\dagger} \widetilde{\mathsf{D}}_{(L)} = \widetilde{\mathsf{D}}_{(L)} \sigma_{L}^{\dagger},~\widetilde{\mathsf{D}}_{(L)} \eta_{(L)} = \sigma_{1}^{\dagger} \widetilde{\mathsf{D}}_{(L)},~\widetilde{\mathsf{D}}_{(L)} \eta_{(L)}^{\dagger} = \sigma_{1} \widetilde{\mathsf{D}}_{(L)}$, the symmetry operator defined in~\eqref{eq:3SP-KW-symmetry-operator} is rewritten as
\begin{equation}
\label{eq:3SP-KW-symmetry-operator-alternative}
    \mathsf{D} = \frac{1}{\sqrt{3}} \left( \frac{1 + \sigma_{1} + \sigma_{1}^{\dagger}}{3} \widetilde{\mathsf{D}}_{(L)} + \frac{1 + \omega^{*}\sigma_{1} + \omega\sigma_{1}^{\dagger}}{3} \widetilde{\mathsf{D}}_{(L)} \sigma_{L} + \frac{1 + \omega\sigma_{1} + \omega^{*}\sigma_{1}^{\dagger}}{3} \widetilde{\mathsf{D}}_{(L)} \sigma_{L}^{\dagger} \right).
\end{equation}
As
\begin{equation}
    \mathsf{U}_{l} \mathsf{U}_{l,l+1} = \left( \begin{array}{ccc}
        \mathsf{U}_{l} &~ \mathsf{U}_{l}\sigma_{l} &~ \mathsf{U}_{l}\sigma_{l}^{\dagger}
    \end{array}\right) \left( \begin{array}{c}
         \frac{1 + \sigma_{l+1} + \sigma_{l+1}^{\dagger}}{3} \cr
         \\
         \frac{1 + \omega^{*}\sigma_{l+1} + \omega\sigma_{l+1}^{\dagger}}{3} \cr
         \\
         \frac{1 + \omega\sigma_{l+1} + \omega^{*}\sigma_{l+1}^{\dagger}}{3}
    \end{array} \right),
\end{equation}
one has
\begin{align}
    \widetilde{\mathsf{D}}_{(L)} =& \left( \begin{array}{ccc}
        \mathsf{U}_{1} &~ \mathsf{U}_{1}\sigma_{1} &~ \mathsf{U}_{1}\sigma_{1}^{\dagger}
    \end{array}\right) \left( \begin{array}{c}
         \frac{1 + \sigma_{2} + \sigma_{2}^{\dagger}}{3} \cr
         \\
         \frac{1 + \omega^{*}\sigma_{2} + \omega\sigma_{2}^{\dagger}}{3} \cr
         \\
         \frac{1 + \omega\sigma_{2} + \omega^{*}\sigma_{2}^{\dagger}}{3}
    \end{array} \right) \cdots \nonumber \\
    &\times \left( \begin{array}{ccc}
        \mathsf{U}_{L-1} &~ \mathsf{U}_{L-1}\sigma_{L-1} &~ \mathsf{U}_{L-1}\sigma_{L-1}^{\dagger}
    \end{array}\right) \left( \begin{array}{c}
         \frac{1 + \sigma_{L} + \sigma_{L}^{\dagger}}{3} \cr
         \\
         \frac{1 + \omega^{*}\sigma_{L} + \omega\sigma_{L}^{\dagger}}{3} \cr
         \\
         \frac{1 + \omega\sigma_{L} + \omega^{*}\sigma_{L}^{\dagger}}{3}
    \end{array} \right) \mathsf{U}_{L}
\end{align}
and
\begin{equation}
    \mathsf{D} = \frac{1}{\sqrt{3}} \mathrm{Tr}_{\text{aux.}} \left( \mathbb{A}_{1} \cdots \mathbb{A}_{L} \right),
\end{equation}
where
\begin{align}
    \mathbb{A}_{l} &= \left( \begin{array}{c}
         \frac{1 + \sigma_{l} + \sigma_{l}^{\dagger}}{3} \cr
         \\
         \frac{1 + \omega^{*}\sigma_{l} + \omega\sigma_{l}^{\dagger}}{3} \cr
         \\
         \frac{1 + \omega\sigma_{l} + \omega^{*}\sigma_{l}^{\dagger}}{3}
    \end{array} \right) \left( \begin{array}{ccc}
        \mathsf{U}_{l} &~ \mathsf{U}_{l}\sigma_{l} &~ \mathsf{U}_{l}\sigma_{l}^{\dagger}
    \end{array}\right) \nonumber \\
    &= \frac{1}{3} \left( \begin{array}{ccc}
        \left( 1 + \sigma_{l} + \sigma_{l}^{\dagger} \right) \mathsf{U}_{l} &~ \left( 1 + \sigma_{l} + \sigma_{l}^{\dagger} \right) \mathsf{U}_{l}\sigma_{l} &~ \left( 1 + \sigma_{l} + \sigma_{l}^{\dagger} \right) \mathsf{U}_{l}\sigma_{l}^{\dagger} \cr
        \\
        \left( 1 + \omega^{*}\sigma_{l} + \omega\sigma_{l}^{\dagger} \right) \mathsf{U}_{l} &~ \left( 1 + \omega^{*}\sigma_{l} + \omega\sigma_{l}^{\dagger} \right) \mathsf{U}_{l}\sigma_{l} &~ \left( 1 + \omega^{*}\sigma_{l} + \omega\sigma_{l}^{\dagger} \right) \mathsf{U}_{l}\sigma_{l}^{\dagger} \cr
        \\
        \left( 1 + \omega\sigma_{l} + \omega^{*}\sigma_{l}^{\dagger} \right) \mathsf{U}_{l} &~ \left( 1 + \omega\sigma_{l} + \omega^{*}\sigma_{l}^{\dagger} \right) \mathsf{U}_{l}\sigma_{l} &~ \left( 1 + \omega\sigma_{l} + \omega^{*}\sigma_{l}^{\dagger} \right) \mathsf{U}_{l}\sigma_{l}^{\dagger}
    \end{array}\right)
\end{align}
is the local tensor of an MPO with periodic boundary condition, for which the physical and auxiliary dimensions are both $3$.





\bibliographystyle{JHEP}
\bibliography{KW.bib}

\end{document}